\documentclass{jfm}
\usepackage{graphicx}
\usepackage{xcolor}
\usepackage{subcaption}
\usepackage[utf8]{inputenc}
\usepackage{newtxtext}
\usepackage{newtxmath}
\usepackage{natbib}
\usepackage{hyperref}
\usepackage{float}
\usepackage{xcolor}
\usepackage{tikz}
\hypersetup{
    colorlinks = true,
    urlcolor   = blue,
    citecolor  = black,
}

\newcommand{\colorline}[1]{\textcolor{#1}{\raisebox{0.5ex}
{\rule{0.4cm}{1.5pt}}}}

\newcommand{\RomanNumeralCaps}[1]
\usepackage{comment}

\usepackage[author={Venkat}]{pdfcomment}

\title {Resolvent analysis to inform viscoelastic coatings for turbulent drag reduction}

\author{Soumen Chakravarty\aff{1} \and
  Venkat Narayanaswamy\aff{1}\corresp{\email{vnaraya3@ncsu.edu}}}
\affiliation{\aff{1}Department of Mechanical and Aerospace Engineering, North Carolina State University, Raleigh, United States}

\begin{document}

\maketitle

\begin{abstract}
\noindent Viscoelastic compliant coatings offer a passive route to modify wall-bounded turbulence; however, their effectiveness for drag reduction remains unresolved. We perform resolvent analysis of turbulent boundary layers over linear viscoelastic continuum, and apply it to incompressible hydrodynamic and compressible aerodynamic zero-pressure-gradient turbulent boundary layers, using both standard and eddy viscosity resolvent formulations. Across a wide range of storage modulus $E$ and coating thickness $H$, viscoelastic surfaces amplify near-wall-cycle-type modes while also attenuating the resolvent gain of very large scale motions (VLSMs) by up to $50\%$, which together result in a reduction of Reynolds stress. For density-matched coatings representative of aqueous incompressible flows, however, these favorable bands lie entirely within the regime where the effective coatings are linearly unstable to traveling wave flutter, rendering them practically unrealizable. Optimizing material damping does not eliminate this but provides a pathway to use weaker sub-optimal interactions. In supersonic flow, the large solid-to-fluid density ratio (O(10³)) shifts the favorable interaction to substantially higher moduli, weakening the achievable reduction in turbulence production to a few percent. However,  the strongest interaction band occurs in the linearly stable regime. These results suggest that compliant wall drag reduction via coupling with high gain modes is fundamentally constrained by flow-induced structural instabilities in incompressible applications, whereas the high density ratios of supersonic flow offer a much narrower but stable window for practical coatings.

\end{abstract}


{\bf MSC Codes }  {\it(Optional)} Please enter your MSC Codes here

\section{Introduction}

Compliant walls have gained significant interest over the last few decades following experiments by \cite{kramer1960boundary}, where it was conjectured that compliant surfaces may delay laminar-to-turbulent transition. Extensive research using stability analysis \citep{landahl1962stability, benjamin1960effects, carpenter1990effect, yeo1988stability, yeo1990hydrodynamic, yeo2001turbulent, tsigklifis2017interaction, pfister2022global} showed that compliant walls can attenuate the growth of Tollmien-Schlichting waves (TSW). However, soft compliant walls are susceptible to transition to turbulence through fluid-induced structural instabilities (FISI). These instabilities grow through the amplification of the elastic deformation modes of the solid, resulting in static divergence (SD) and traveling wave flutter instabilities (TWF) that were observed experimentally in prior works \citep{gad1984interaction,gad1986response}. Fluid-induced structural instabilities have also been demonstrated in compressible flows in the laminar \citep{deka2026stability} and turbulent boundary layers \citep{chakravarty2026unstable}. While the physics of TSW attenuation and amplification of FISI are well understood, the interaction of viscoelastic surfaces with turbulent boundary layers, and its effect on turbulent skin friction remain relatively underexplored.

Early works on turbulent drag reduction using compliant walls focused on the modification of turbulent bursts using resonant compliant surfaces \citep{bushnell1977effect}. One of the first experimental evidence on drag reduction with viscoelastic coatings was provided by \cite{lee1993investigation}. The authors also observed that drag reduction was accompanied by an increase in the near-wall streak spacing and an upward shift of the log layer; these features have also been observed for flows over drag- reducing riblets \citep{suzuki1994turbulent, choi1993direct,duan2014direct,garcia2011hydrodynamic} and anisotropic permeable walls \citep{gomez2017turbulent,abderrahaman2017analysis}. \cite{choi1997turbulent} tested stiff silicone-based compliant surfaces and obtained a drag reduction of 7\%; concomitantly, a reduction in turbulent fluctuations was also observed. Recent experiments \citep{zhang2017deformation, wang2020interaction, huynh2021experiments, greidanus2022response, lu2024scaling, lu2025analysis} have probed the turbulent flow-structure interactions in much greater detail. Detailed measurements of wall deformations and turbulent flow-fields by \cite{zhang2017deformation} and \cite{wang2020interaction} showed that the deformations of the viscoelastic compliant walls are driven by pressure fluctuations in the log-layer, with the deformation wavelengths scaling strongly with the compliant layer thickness. \cite{greidanus2022response} observed turbulence-driven deformations for low bulk velocities, which gradually transitioned to spanwise-uniform modes resembling TWF with increasing flow-speeds. However, all of the above studies reported an increase in turbulent stresses and overall drag across all flow speeds considered.  

Direct numerical simulations (DNS) by \cite{xu2003turbulence} reported very little modification of mean turbulent statistics using compliant walls. It was concluded that the natural phase relation between wall pressure and wall-normal velocity over the compliant wall is unfavorable for counteracting the ejections and bursts associated with near-wall turbulence. Similar efforts by \cite{kim2014space} also reported little modification of turbulent statistics for stiff compliant walls, while softer walls led to the growth of large-amplitude traveling waves that increased the drag. Recent DNS studies of turbulent flows over hyper-elastic walls \citep{rosti2017numerical, esteghamatian2022spatiotemporal} showed that the compliant surfaces preferentially formed spanwise-coherent waves with increasing softness, resulting in increased turbulent fluctuations and drag. \cite{koseki2025understanding} showed that wall elasticity promotes increased wall-normal fluctuations through strong near-wall ejections, which are absent in the case of static rough walls with similar topologies. 

%Thus, apart from early experimental successes \citep{lee1993investigation,choi1997turbulent}, consistent demonstration of drag reduction potential of compliant walls in incompressible aqueous flows have not yet been established. In the supersonic regime, recent experiments \cite{chakravarty2024investigations} have demonstrated concomitant reduction in skin friction and Reynolds stresses for a Mach 2.5 turbulent boundary layer over a viscoelastic surface. However, lack of time-resolved near-wall data hindered comprehensive understanding of the interaction. 

Given the multitude of governing parameters in flow-compliant wall interactions, extensive parametric investigations at high Reynolds numbers using detailed computations seem impractical. Initial lower-order models developed by \cite{duncan1986response} modeled the unsteady deformations of a viscoelastic wall due to an assumed pressure pulse from a turbulent burst. Application of resolvent analysis \citep{mckeon2010critical} on wall- bounded turbulent flows has enabled the study of turbulence control from an input-output perspective. Resolvent analysis has been successfully applied to model different classes of wall modification such as riblets, permeable walls and compliant membranes \citep{chavarin2020resolvent, chavarin2021resolvent, luhar2015framework, luhar2016design, song2026structured}. Using resolvent analysis of turbulent channel flow over spring-backed compliant walls, \cite{luhar2015framework} showed that certain compliant wall designs hold potential to suppress large-scale turbulent motions; however, these walls lead to amplification of spanwise-uniform modes similar to \cite{kim2014space}. The framework was extended by adding stiffness, flexure and anisotropy \citep{luhar2016design}. The compliant walls exhibited sharp transitions of mode attenuation/amplification behavior  across resonant frequencies, but the amplification of detrimental two-dimensional modes persisted. Recent efforts by \cite{liu2021structured} preserved part of the previously uncertain nonlinear terms $u.\nabla u$ using a simplified block-diagonal form to frame the problem into structured singular value optimization. \cite{song2026structured} extended this framework to turbulent flows over spring backed compliant surfaces, demonstrating efficacy of compliant surfaces to attenuate both near-wall streamwise vortices and large-scale motions. However, these optimal designs were still shown to amplify two-dimensional modes significantly. 

The coherent structure based approach \citep{luhar2015framework, luhar2016design,  song2026structured} provides a better understanding of how energetically relevant turbulent scales may be modified by the wall admittance due to a mass spring-damper system. However, there is still a lack of understanding of how practical viscoelastic layers may be designed to favorably interact with turbulent boundary layers. Viscoelastic continuum exhibits multiple modes, thickness-dependent behavior, supports the propagation of elastic waves and in-plane deformations, all of which are not captured by spring-damper models. For example, whereas the compliant wall layer thickness has been shown to be of paramount importance \citep{zhang2017deformation, wang2020interaction}, so far it remains unexplored in the input-output framework. While preliminary works using resolvent analysis of turbulent flows over viscoelastic layers have been undertaken \citep{chakravarty2024analysis, bhagwat2025towards} over isotropic and anisotropic coatings, a detailed analysis over different flow and solid regimes remains unexplored. In this work, we extend the resolvent framework proposed by \cite{luhar2015framework} to linear isotropic viscoelastic materials interacting with subsonic and supersonic turbulent boundary layers, while exploring all solid parameters extensively. We investigate the effect of viscoelastic layers on modulating energetic coherent structures in turbulent boundary layers and derive realistic margins of operation using complementary linear stability analysis.

\section{Methodology}

\subsection{ Problem description}

The physical problem consists of a fluid and a solid domain. The fluid domain comprises a zero pressure gradient turbulent boundary layer, with the $x_1, x_2$ and $x_3$ axes representing the streamwise, wall-normal and spanwise directions, respectively. The fluid domain is semi-bounded in the $x_2$ direction, starting at the wall ($x_2 = 0$). A bounded viscoelastic coating lies beneath the fluid domain, extending from $x_2 = 0$ to $x_2 = -H$, where $H$ is the thickness of the viscoelastic patch. The viscoelastic coating is fixed to a rigid surface at $x_2 = -H$. The domain is unbounded in the $x_1$ and $x_3$ directions. In this study, both incompressible and compressible turbulent boundary layers are individually considered. In the steady equilibrium condition, the compliant surface is assumed to be undeformed in the $x_2$ direction, such that the steady equilibrium interface lies at $x_2 = 0$. Due to the undeformed state of the compliant surface along $x_2$, the turbulent boundary layer profile considered in the steady equilibrium will be identical to the turbulent mean profile over a rigid wall. The viscoelastic solid and fluid are coupled with each other through the constraints imposed at the interface $x_2 = 0$, which includes continuity of surface traction and velocities.        

\begin{figure}
    \centering
    \includegraphics[scale = 0.35]{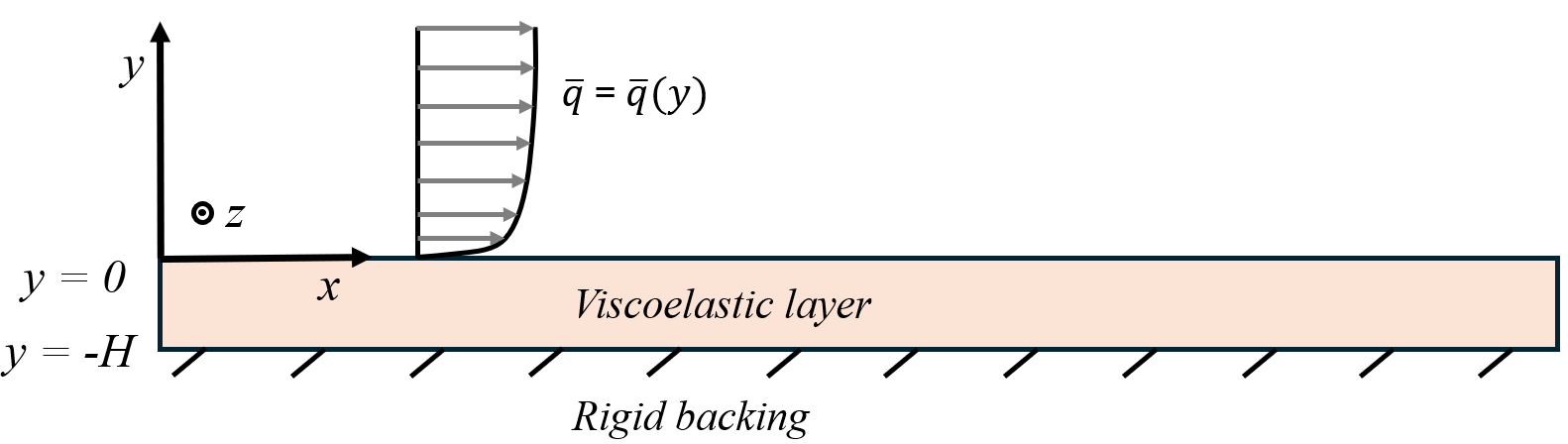}
    \caption{Schematic of a turbulent boundary layer over a viscoelastic implant.}
    \label{fig:Figure 1}
\end{figure}
 
\subsection{Governing equations and linearization}

 The mean-subtracted compressible Navier-Stokes equations are given by

\begin{align}
&\frac{\partial \rho}{\partial t}
+ \nabla \cdot (\rho u) = 0 \label{eq:1}\\
%\tag{1}
%\\[1em]
&\rho \frac{D u_i}{D t}
+ \frac{1}{\gamma Ma^2}\nabla p
-\frac{1}{Re}
\left[
\lambda \nabla^2 u_i
+\mu \frac{\partial}{\partial x_i}(\nabla \cdot u)
+(\nabla \mu)\cdot(\nabla u_i)
+(\nabla \mu)\frac{\partial u_j}{\partial x_i} +\frac{\partial}{\partial x_j}(\lambda \nabla \cdot u)
\right] = 0, \label{eq:2}\\
%\tag{2}
%\\[1em]
&\rho \frac{D T}{D t}
-(1-\gamma)p\nabla\cdot u
-\frac{\gamma}{Re}
\nabla\cdot
\left(
\frac{\mu}{Pr}\nabla T
\right)
-\lambda (\nabla\cdot u)^2
-\frac{\mu}{2}
\left(
\nabla u + (\nabla u)^T
\right)^2 = 0
\label{eq:3}
\end{align}

All quantities are normalized by their respective freestream values, such that the Reynolds number is given by $Re = \rho_\infty U_\infty \delta/\mu_\infty$, and Mach number $Ma = U_\infty/\sqrt{\gamma R T_\infty} $. Here, $\delta$ is the boundary layer thickness, $\gamma$ is the ratio of specific heats, and $R$ is the universal gas constant. The Prandtl number is defined as $Pr = \mu_\infty c_p/\kappa_\infty$, where $c_p$ is the specific heat and $\kappa$ is the thermal conductivity of the fluid. We assume the fluid to be thermally perfect, with constant values of $\gamma = 1.4$ and $Pr = 0.72$ in this study. Here, $\mu$ and $\lambda$ are the first and second coefficients of viscosity. Invoking the Stokes assumption, we obtain $\lambda = -2\mu/3$. The molecular viscosity $\mu$ is temperature dependent and obeys Sutherland's law. Additionally, the ideal gas equation is used to close [\ref{eq:1} - \ref{eq:3}], and is given by $p = \rho T$. 
We shall also use the incompressible form of Navier-Stokes equations, which are given by

\begin{align}
&\frac{D u_i}{D t}
+ \nabla p
-\frac{1}{Re}\nabla^2 u_i= 0
\label{eq:4}
\\
&\nabla.u = 0
\label{eq:5}
\end{align}
The governing equations are thus expressed in the form 
\begin{equation}
    \frac{\partial \textbf{\textit{q}}_f}{\partial t} = N(\textbf{\textit{q}}_f) 
    \label{eq:6}
\end{equation}
Here $\textbf{\textit{q}}_f = [u_f,v_f,w_f,\rho_f,T_f]^T \in \mathbb{R}^{5n_f}$ are the state variables of the fluid for compressible flow, and $n_f$ is the size of the domain. For incompressible flow, we have $\textbf{\textit{q}}_f = [u_f,v_f,w_f,p_f]^T \in \mathbb{R}^{4n_f}$. For brevity, we shall only show the compressible form of equations in the subsequent formulation. Using the parallel flow assumption, the state variables are decomposed into statistically stationary mean and fluctuating components, given by
\begin{equation}
    \textbf{\textit{q}}(x,y,z,t) = \bar{\textbf{\textit{q}}}(y)+\textbf{\textit{q}}^{\prime}(x,y,z,t)
    \label{eq:7}
\end{equation}
Here, $ \bar{\textbf{\textit{q}}}(y) = [\bar{u}_f,0,0,\bar{\rho}_f,\bar{T}_f]^T$ is the mean compressible turbulent boundary layer profile. The pressure is assumed to be constant across the boundary layer, such that $\bar{p} = 1$. The equation of state therefore becomes $\bar{\rho}\bar{T} = 1$. On linearizing this equation of state, we obtain $p = \bar{\rho}T + \bar{T}\rho$.  The linearized Navier-Stokes equation for perturbations about the turbulent mean profile is given by
\begin{align}
    \frac{d\textbf{\textit{q}}_f'}{dt}  = L_f\textbf{\textit{q}}_f' +\textbf{\textit{f}}'
    \label{eq:8}
\end{align}
Here $L_f$ represents the linearized Navier-Stokes operator, and $\textbf{\textit{f}}'$ is the forcing which includes the nonlinear terms from the Navier-Stokes equations and any external forcing. 

In order to define the viscoelastic solid, we used the following quantities, which are non-dimensionalized by the freestream fluid quantities. 

\begin{align}
    \rho_s = \frac{\rho^*_s}{\rho_\infty}, \quad h = \frac{h^*}{\delta}, \quad \mu_s = \frac{\mu^*_s}{\rho_\infty U_\infty \delta}, \quad C_t = \frac{C^*_t}{U_\infty}
    \label{eq:9}
\end{align}
Here, $\rho^*_s$ is the solid density, $h^*$ is the thickness of the solid, and $C^*_t$ is the material shear speed. The viscoelastic damping is modeled using the Kelvin-Voigt model, and $\mu^*_s$ is the solid viscosity or damping coefficient. To model the viscoelastic solid, we use the linearized elasticity equation, which is given by
\begin{align}
    \rho_s \frac{\partial u'_s}{\partial t} - \nabla.\sigma_s(\eta',u'_s) = 0
    \label{eq:10}
\end{align}
Here, $\sigma_s$ is the infinitesimal stress tensor, $u'_s$ is the solid velocity perturbation and $\eta'$ is the solid displacement perturbation. The stress tensor is given by
\begin{align}
    \sigma_s = G\left( \nabla \eta' + (\nabla \eta')^T\right) + \left(K - \frac{2G}{3}\right)\nabla.\eta' \mathbf{I}
    \label{eq:11}
\end{align}
Here, $G = \rho_sC_t^2 - i\omega\mu_s$ is the complex shear modulus, and $K$ is the bulk modulus of the solid. Combining [\ref{eq:10}] and [\ref{eq:11}], the final governing equation for small perturbations in the solid is given by the Navier's equation \citep{yeo2001turbulent}
\begin{align}
    \rho_s \frac{\partial u'_s}{\partial t} &= G\nabla^2\eta + \left(K + G/3\right)\nabla(\nabla.\eta)  \label{eq:12}\\
    \frac{\partial \eta'}{\partial t} &= u'_s
    \label{eq:13}
\end{align}
The coupled system of equations can be given by
\begin{align}
    \begin{pmatrix}I_f & 0 \\0 & I_s\end{pmatrix}\frac{\partial}{\partial t}\begin{pmatrix}q_f' \\q_s'\end{pmatrix}-\begin{pmatrix}L_f(\bar{q}) & C_{fs} \\C_{sf} & L_s\end{pmatrix}\begin{pmatrix}q_f' \\q_s'\end{pmatrix}=\begin{pmatrix}M_f & 0  \\0 & 0\end{pmatrix}\begin{pmatrix}f_f' \\f_s'\end{pmatrix}
    \label{eq:14}
\end{align}
Here, $q'_s = [u'_s, \eta']^T$ are the state variables of the solid. The diagonal block matrices $L_f$ and $L_s$ are the linearized governing equations for the fluid [\ref{eq:1} -- \ref{eq:3}] and the solid [\ref{eq:12} -- \ref{eq:13}] respectively. The off-diagonal matrices $C_{sf}$ and $C_{fs}$ perform two-way coupling of the fluid and solid at the interface $y = 0$, using the interface conditions. We do not assume any internal sources of forcing within the solid, hence the forcing term is zero for the solid part of equations. In the fluid part, forcing is applied using the mask $M_f$, which is used in the application of boundary and interface conditions, and can also be used to spatially constrain the forcing.  

The solid and fluid equations are coupled through the continuity of velocity and stresses across the interface. The terms in $C_{fs}$ are used to impose continuity of velocity across the interface, which are given by
\begin{align}
    u_s = u_f + \eta_{2, y = 0} \frac{\partial \bar{U}}{\partial y}, \quad v_s = v_f, \quad w_s = w_f
    \label{eq:15}
\end{align}
The continuity of stresses is managed by the operator $C_{sf}$, and the equations are given by
\begin{align}
    &\frac{\bar{\mu}_w}{Re}\left( \frac{\partial u'_f}{\partial y} + \frac{\partial v'_f}{\partial x} + \eta'_2 \frac{\partial^2 \bar{U}_f}{\partial y^2}\right) = G\left( \frac{\partial \eta'_1}{\partial y} + \frac{\partial \eta'_2}{\partial x}  \right),  \label{eq:16}\\
    & -p + \frac{2\bar{\mu}_w}{Re} \frac{\partial v'_f}{\partial y} + \bar{\lambda}\nabla. u = 2G \frac{\partial n'_2}{\partial y} + \left(K + \frac{G}{3}\right)\nabla. \eta,  \label{eq:17}
     \\
    &\frac{\bar{\mu}_w}{Re}\left( \frac{\partial w'_f}{\partial y} + \frac{\partial v'_f}{\partial z} \right) = G\left( \frac{\partial \eta'_3}{\partial y} + \frac{\partial \eta'_2}{\partial z}  \right) 
    \label{eq:18}
\end{align}
The fluid temperature fluctuations are assumed to be zero ($T_f' = 0$) on the surface of solid, and the wall is assumed to be adiabatic. No boundary condition is applied to the fluid density fluctuations $\rho'_f$ at $y = 0$. For the baseline rigid wall case, velocity fluctuations are set to zero at the wall. For $y \rightarrow\infty$, we use the characteristic boundary conditions \citep{thompson1987time} with a sponge zone (5\% of total wall-normal extent at the top boundary) which ensures no non-physical waves enter the domain. In the case of incompressible flows, at $y \rightarrow \infty$, all the state variables can be set to zero. 
Since the solid is fixed to a rigid surface at $y = -h$, the boundary conditions at $y = -h$ for the solid reads ($\eta', u'_s = 0$). 

Since the streamwise and spanwise directions are homogeneous, the fluctuations of state variables and forcing can be expressed in terms of their Fourier-transformed coefficients as 
\begin{align}
    \textbf{\textit{q}}'(x,y,z,t) &= \int_{-\infty}^{\infty} \int_{-\infty}^{\infty} \int_{-\infty}^{\infty} \hat{\textbf{\textit{q}}}(y)e^{i(k_xx+k_zz-\omega t)} dk_x dk_z d\omega  \label{eq:19}\\
    \textbf{\textit{f}}'(x,y,z,t) &= \int_{-\infty}^{\infty} \int_{-\infty}^{\infty} \int_{-\infty}^{\infty} \hat{\textbf{\textit{f}}}(y)e^{i(k_xx+k_zz-\omega t)} dk_x dk_z d\omega    
    \label{eq:20}
\end{align}
Here, $\textbf{k} = (k_x, k_z,\omega)$ is the wavenumber-frequency triplet. The phase speed of the mode is given by $c = \omega/k_x$. The inner scaled terms will be defined using the superscript $+$, such as $c^+$ for mode speed. On Fourier transforming [\ref{eq:14}] and writing in terms of its Fourier coefficients, we get
\begin{align}
    \left[-i\omega I-\begin{pmatrix}L_f(\bar{q}) & C_{fs} \\C_{sf} & L_s\end{pmatrix}\right]\begin{pmatrix}\hat{q}_f' \\\hat{q}_s'\end{pmatrix}=\begin{pmatrix}M_f \hat{f}' \\0\end{pmatrix}
    \label{eq:21}
\end{align}
Using $\hat{q}_\textbf{k} = [\hat{q}_f, \hat{q}_s]^T_\textbf{k} $ to denote the full state corresponding to the triplet $\textbf{k}$,
\begin{equation}
    \hat{\textbf{\textit{q}}}_{\textbf{k}} =  (-i\omega I - L_{\textbf{k}})^{-1} M\hat{\textbf{\textit{f}}}_{\textbf{k}} \implies \hat{\textbf{\textit{q}}}_{\textbf{k}} =  H_{\textbf{k}}\hat{\textbf{\textit{f}}}_{\textbf{k}}
    \label{eq:22}
\end{equation}
Here, $H_{\textbf{k}}$ is known as the resolvent operator, which is the transfer function between the forcing $\textbf{\textit{f}}_k$ and the flow/solid response $\textbf{\textit{q}}_k$. An SVD of the resolvent operator $H_k$ provides a set of orthonormal forcing $\textbf{\textit{f}}_{k,m}$ and response modes $\textbf{\textit{q}}_{k,m}$ under an $L_2$ norm. This norm is enforced using appropriate weighting matrices such that
\begin{align}
    W_q \hat{\textbf{\textit{q}}}_{\textbf{k}} = \left(W_q H_k W^{-1}_f\right) W_f \hat{\textbf{\textit{f}}}_{\textbf{k}} \implies W_q \hat{\textbf{\textit{q}}}_{\textbf{k}} = H^s_{\textbf{k}} (W_f\hat{\textbf{\textit{f}}}_{\textbf{k}}) 
    \label{eq:23}
\end{align}
The SVD of the scaled resolvent operator $H^s_{\textbf{k}}$ is given by
\begin{align}
    H^s_{\textbf{k}} = \Psi_{\textbf{k}}\Sigma_{\textbf{k}}\phi^*_{\textbf{k}},
    \label{eq:24}
\end{align}
where $\Psi_{\textbf{k}} = [\Psi_{{\textbf{k}},1}, \Psi_{{\textbf{k}},2},..., \Psi_{{\textbf{k}},n}]$ is the set of left singular vectors which forms an orthonormal basis for the response and $\phi_{\textbf{k}} = [\phi_{{\textbf{k}},1},\phi_{{\textbf{k}},2},..., \phi_{{\textbf{k}},n}]$ is the set of right singular vectors which forms an orthonormal basis for the forcing, where the superscript $^*$ denotes the Hermitian transpose. The unscaled forcing and response modes can be recovered as $\hat{\textbf{\textit{q}}}_{{\textbf{k}},m} = W_q^{-1}\Psi_{{\textbf{k}},m}$ and $\hat{\textbf{\textit{f}}}_{{\textbf{k}},m} = W_f^{-1}\phi_{{\textbf{k}},m}$. The matrix $\Sigma_{{\textbf{k}}} =$ diag$(\sigma_{{\textbf{k}},1}, \sigma_{{\textbf{k}},2},...,\sigma_{{\textbf{k}},n})$ are the singular values which represent the amplification (gain) for a forcing-response pair. The singular values are in descending order $(\sigma_{{\textbf{k}},1}\geq\sigma_{{\textbf{k}},2}\geq...\geq\sigma_{{\textbf{k}},n})$, and the first singular value is known as leading resolvent gain. If the resolvent operator is low rank, i.e. $\sigma_{{\textbf{k}},1}>>\sigma_{{\textbf{k}},2}$, then the scaled resolvent operator $H^s_{\textbf{k}}$ can be expressed as
$\sigma_{{\textbf{k}},1}\Psi_{{\textbf{k}},1} \approx H^s_{\textbf{k}}\phi_{{\textbf{k}},1}$. 

We seek to maximize the response only within the fluid domain. Thus, in our choice of norm, we do not use any weights for the solid state variables. For the forcing norm, we only choose to force the momentum equations, thus $W_f = \text{diag}[I,I,I,0,0]$. For the response, we use the Chu norm $W_r = \text{diag}\left[\bar{\rho},\bar{\rho},\bar{\rho},\frac{R \bar{T}}{\bar{\rho}} , \frac{R \bar{\rho}}{(\gamma-1) \bar{T}}\right]$ for compressible flow and the kinetic energy norm $W_r = \text{diag}[I,I,I,0]$ for incompressible flow. The block matrices $W_r$ and $W_f$ are constructed using the norm coefficients and numerical quadrature weights which depend on the discretization scheme and the wall-normal grid.

For linear stability analysis, the forcing $\hat{f}'$ is absent in [\ref{eq:21}]. In this case, the eigenvalue form is recovered
\begin{align}
    -i\omega\hat{q}' = L\hat{q}' 
    \label{eq:25}
\end{align}
Here, $\omega = \omega_r + i\omega_i$ is the complex eigenvalue. The coupled system becomes convectively unstable when the least stable eigenvalue lies in the upper half of the complex plane.

\subsection{Eddy viscosity model}

In traditional resolvent analysis, all the nonlinear effects are included in the forcing term, which is also assumed to be uniform across frequency. Recent developments \citep{illingworth2018estimating, symon2021energy, symon2023use, morra2019relevance,fan2024eddy} include an eddy viscosity, which models the effects of turbulent diffusion and dissipation \citep{symon2023use}, resulting in better agreement of resolvent modes with DNS profiles for both incompressible \citep{symon2021energy, symon2023use, morra2019relevance, illingworth2018estimating} and compressible turbulent boundary layers \citep{fan2024eddy}. Thus, we will also consider the effect of eddy viscosity based resolvent modes on fluid-structure coupling. The eddy viscosity is added to the viscosity dependent terms in the momentum and energy equations as 

\begin{align}
    \mu = \mu +\mu_t, \quad \lambda = \lambda+\lambda_t, \quad k = \frac{\mu}{Pr} + \frac{\mu_t}{Pr_t}
    \label{eq:26}
\end{align}
Here, $Pr_t = 0.85$ \citep{zhang2014generalized} is the turbulent Prandtl number which is invariant to Mach number and wall temperature conditions \citep{zhang2018direct}, as well as remains constant across the boundary layer. For the compressible boundary layer profile, the eddy viscosity profile is obtained using the mixing length model, which is defined as
\begin{align}
    \mu_t = \bar{\rho}\ell^2_m \frac{d\bar{U}}{dy}, \quad \ell_m = \begin{cases} \kappa y \left[1 - \exp\left(\dfrac{-y^*}{A^+ + f(M_\tau)}\right)\right]^2, & y \leq \delta_c \\[10pt] C_\mu \delta, & y > \delta_c \end{cases}
    \label{eq:27}
\end{align}
Here, $\kappa = 0.41$, $\bar{\mu}$ is the molecular viscosity, and $y^* = y/\delta^*_v$ is the semi-local wall-normal coordinate, where $\delta_v^* = {\bar{\mu}}/{\bar{\rho}u^*_\tau}$, $f(M_\tau) = 19.3 M_\tau$, $A^+ = 17$, $C_\mu = 0.09$ and $\delta$ is the boundary layer thickness. In the absence of thermodynamic property variations, the incompressible form of eddy viscosity is recovered. The crossover distance $\delta_c$ is the location where values of both the expressions are equal. The inner and outer expressions are blended using the following form \citep{johnson1985mathematically}
\begin{align}
    \mu_t = \mu_{t,\text{o}}\left(1 - \exp(-\mu_{t,i}/\mu_{t,o})\right)
    \label{eq:28}
\end{align}
\subsection{Numerical implementation}

For compressible flow, the fluid domain was discretized into $n_f = $ 401 points in the wall-normal $(y)$ direction using Chebyshev collocation points. For incompressible flow, $n_f = 201$ points are used. A grid transformation was used to maintain near-wall resolution, given by $y_n = ay/(b-y)$ \citep{schmid2002stability}, where $-1\leq y \leq 1$ represents the domain for Chebyshev polynomials. The domain size was kept to be $y_{max} = 4\delta$. Here, $a = \frac{y_i y_{max}}{y_{max}-2y_{i}}$ and $b = 1+\frac{2a}{y_{max}}$. The parameter $y_i$ is the wall-normal location below which $50\%$ of the points are present. For the compliant wall, $n_s = $ 41 Chebyshev points were used to discretize the wall from $y = 0$ to $y = -H$. For subsonic flows, the dataset at $Re_\tau = 1990$ \citep{sillero2013one,sillero2014two} was used. The supersonic turbulent boundary layer dataset by \cite{pirozzoli2011turbulence} at $M_\infty = 2$ and $Re_\tau = 1110$ was used to obtain the mean profiles for the compressible cases.

\begin{figure}
    \centering
    \includegraphics[width=0.7\linewidth]{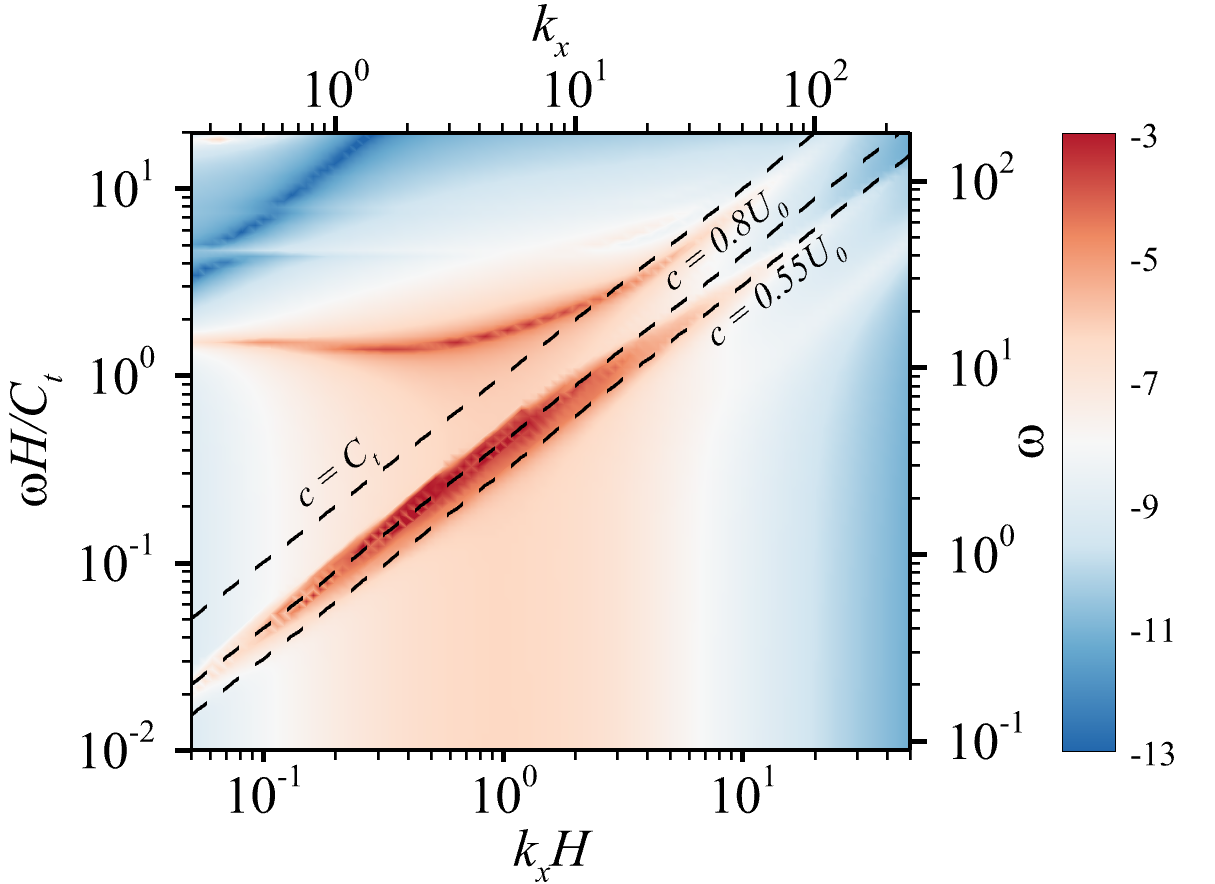}
    \caption{PSD of wall-normal deformations of viscoelastic surface for $Re_\tau = 6700$ from \cite{lu2024scaling}.}
    \label{fig:Figure 2}
\end{figure}

\section{Validation}

Experimental results \citep{zhang2017deformation, wang2020interaction, lu2024scaling} have shown that viscoelastic surface deformations in the stable regime are driven by pressure fluctuations in the log-layer. The wall deformations are known to advect at speeds between $0.55 U_0 \text{ and } 0.9U_0$ across different experiments, which is concomitant with the broad range of convection speeds for pressure fluctuations in the log-layer. For a particular wavenumber-frequency triplet $\textbf{k}$, the PSD of wall-normal deformations can be extracted from the coupled resolvent operator using the procedure described by \cite{liu2020input}, assuming a spatio-temporally uncorrelated forcing structure. We used the experimentally obtained mean turbulent boundary layer profile by \cite{lu2024scaling} at $Re_\tau = 6700$ to compute the wall deformation PSD. For the particular case described by \cite{lu2024scaling}, the solid parameters are given by $E = 4.5, H = 0.1, \rho_s = 0.5, \text{and} \tan(\phi) = 0.01$. Figure \ref{fig:Figure 2} presents the power spectral density (PSD) of the wall-normal surface deformations for spanwise uniform ($k_z = 0$) modes. Two clear advection bands can be observed at speeds $c = 0.55U_0$ and $c = 0.8U_0$, which agree strongly with the range of convection velocities observed in the experiments. The slower band occurs for higher wavenumbers, whereas the faster band occurs for smaller wavenumbers. The top-most band showing strong response represents elastic modes of the compliant wall, which includes the advection band for Rayleigh waves. These waves propagate at the material shear speed $c = C_t$. For this particular case, $C_t = 1.8 U_0$, which places the critical layer outside of the boundary layer edge. \cite{wang2020interaction} also consistently observed deformation peaks at $k_x H = 2\pi/3$; such a distinct peak is not observed in this case.

%\begin{equation}
%    \hat{\textbf{\textit{q}}}_{\textbf{k},\eta_2} = C_{\eta_2}H_\textbf{k}\hat{\textbf{\textit{f}}}_{\textbf{k}} = G_{\eta_2, \textbf{k}}\hat{\textbf{\textit{f}}}_{\textbf{k}}
%\end{equation}
%Here, $C_{\eta_2}$ is the output operator to extract the wall-normal surface displacement from the full state.  

\section{Results and Discussion}

\subsection{Rank of Resolvent Operator}

\begin{figure}
    \centering
    \includegraphics[width=1\linewidth]{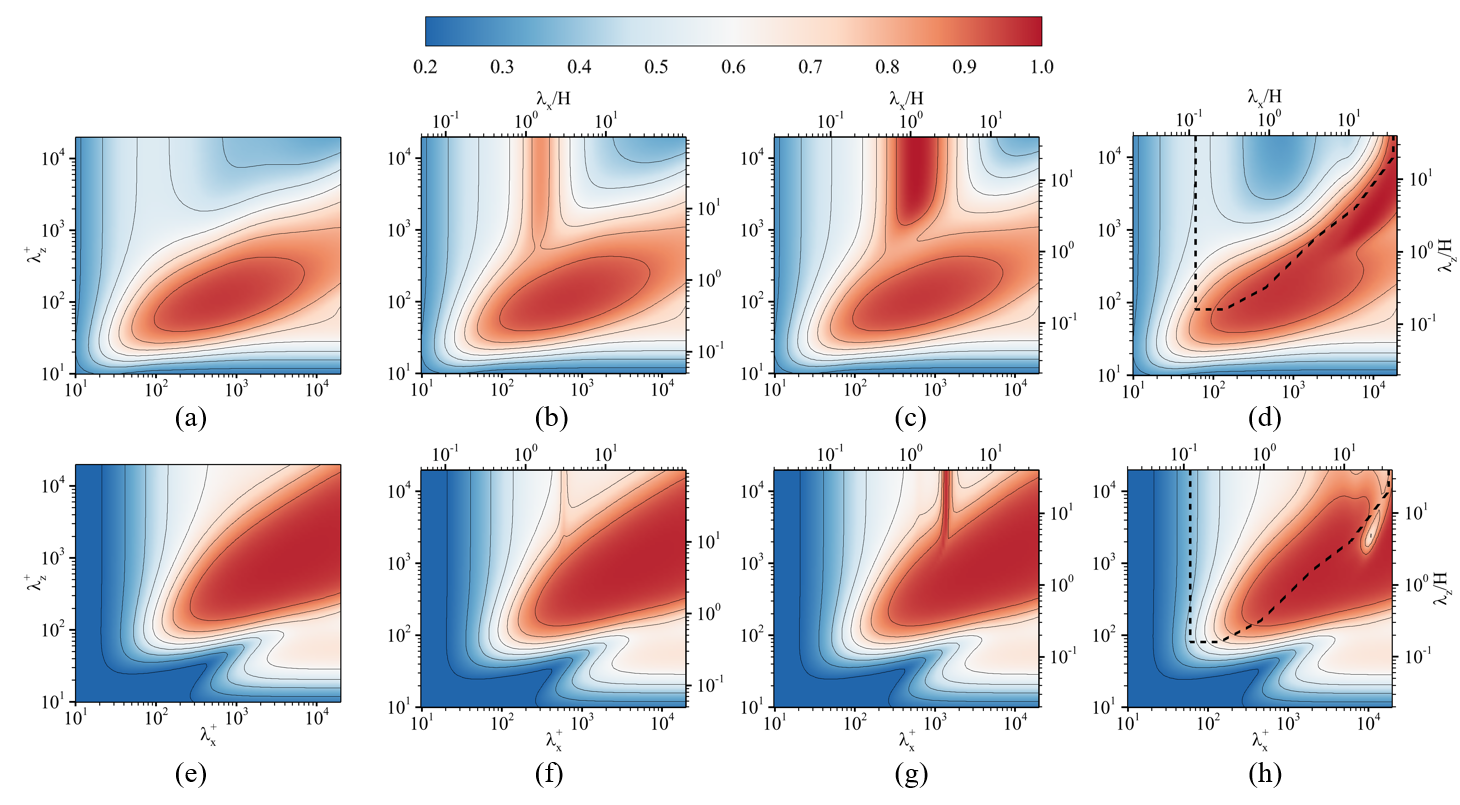}
    \caption{Fraction of energy captured by the first resolvent mode $\sigma^2_1/\sum_{i}\sigma_i^2$ across different wavelength combinations $(\lambda^+_x, \lambda^+_z)$ at $c^+ = 10$ (a-d) and $c^+ = 16$(e-h)  for an incompressible turbulent boundary layer at $Re_\tau = 1990$. The cases presented are \emph{(a,e)} Rigid wall, \emph{(b,f)} viscoelastic surface with $E = 0.5, H = 0.1$, \emph{(c,g)} viscoelastic surface with $E = 0.5, H = 0.25$, and  \emph{(d,h)} viscoelastic surface with $E = 0.01, H = 0.25$. The solid viscosity and density are maintained constant as $\mu_s = 0.005$ and $\rho_s = 1$.  } 
\label{fig:Figure 3}
\end{figure}

In this section, we discuss the low-rankedness of the incompressible resolvent operator in the presence of fluid-structure coupling. The rank of the resolvent operator is defined by the fraction of energy contained in the leading resolvent mode, and can be expressed as $R = \sigma^2_1/\sum_i \sigma^2_i$. As discussed in the previous section, if the resolvent operator is strongly low rank for a given wavenumber frequency combination $\textbf{k}$, the leading resolvent mode can be conveniently used to study the flow characteristics. Figures \ref{fig:Figure 3}(a-d) show the low rank map of the incompressible resolvent operator at $Re_\tau = 1990$ for a range of inner scaled wavelengths ($\lambda^+_x, \lambda^+_z$) and fixed wave speed $c^+ = c/u_\tau = 10$, for walls with varying compliance properties. Near-wall turbulent structures have primarily been shown to convect at these speeds \citep{kim1993propagation, del2009estimation}. Figures \ref{fig:Figure 3}(e-h) provide similar information, but for modes located further from the wall, with wave speed of $c^+ = 16$. Figures \ref{fig:Figure 3}a and \ref{fig:Figure 3}d show the low rank spectrum for the canonical rigid wall. In Figure \ref{fig:Figure 3}a, the resolvent operator is strongly low rank ($R > 0.9$) for the majority of intermediate to larger streamwise scales and smaller to intermediate spanwise scales, showing a strong degree of anisotropy $(\lambda_x > \lambda_z)$. These observations are identical to the discussions by \cite{mckeon2010critical} and \cite{moarref2013model}. \cite{moarref2013model} also showed that regions of strong low rank behavior coincide well with the streamwise turbulent energy spectra from DNS. In the case of Figure \ref{fig:Figure 3}a, low rank spectrum peaks at $(\lambda^+_x,\lambda^+_z) = (800-1000, 100)$, which coincides well with the near-wall cycle in wall turbulence \citep{robinson1991coherent}. In the subsequent sections, the triplet $(\lambda_x^+,\lambda_z^+,c^+)= (1000,100,10)$ will be used to represent the near-wall cycle. With the introduction of a viscoelastic wall with $E = 0.5, H = 0.1$ in Figure \ref{fig:Figure 3}b, the spectrum exhibits an additional vertical low rank region centered at $\lambda^+_x = 320$ and spanning intermediate to large spanwise wavelengths. The vertical band, however does not affect the canonical low rank region introduced in Figure \ref{fig:Figure 3}a. On increasing the thickness of the viscoelastic coating to $H = 0.25$, as shown in Figure \ref{fig:Figure 3}c, the vertical low rank region grows stronger and covers a larger range of streamwise wavelengths, and also shifts toward higher wavelength $\lambda^+_x = 630$. However, the canonical low rank region pertaining to the near-wall cycle remains unchanged. When scaled by coating thickness $H$, the wavelengths of the compliance-induced low rank regions for the cases described in Figure \ref{fig:Figure 3} (b-c) are given by $\lambda_x/H = 1.6$ and $\lambda_x/H = 1.3$ respectively, suggesting a strong influence of the coating thickness on the selection of scales that participate in fluid-structure coupling. In Figure \ref{fig:Figure 3}d, the stiffness is reduced to $E = 0.01$ such that the viscoelastic coating becomes unstable for a large region in the spectrum. This unstable region lies above the black dashed line. Due to the instability, amplification of the fluid-structure modes along the vertical bands disappear completely, and a new strong low rank region is found for large ($\lambda^+_x, \lambda^+_z$). The canonical low rank region related to the near-wall cycle is found to remain unaffected, similar to previous configurations. 

We move to turbulent scales moving at higher speeds of $c^+ = 16$, which corresponds to $c = 0.6$ for the current Reynolds number. Large scale turbulent structures have scale dependent convection velocities which lie approximately in the range $0.6 U_\infty - 0.7 U_\infty$. The low rank map for the canonical rigid wall configuration is shown in Figure \ref{fig:Figure 3}e. Compared to Figure \ref{fig:Figure 3}a, the resolvent gain becomes low rank for much larger scales, however, the strongly low-rank combinations remain anisotropic ($\lambda_x>\lambda_z$). For the viscoelastic case with $E = 0.5, H = 0.1$ shown in Figure \ref{fig:Figure 3}f, a narrow vertical low rank region is observed, although with a much weaker intensity compared to the corresponding near-wall case (Figure \ref{fig:Figure 3}b). In Figure \ref{fig:Figure 3}g for $E = 0.5, H = 0.25$, the intensity of compliance-induced effects increases compared to Figure \ref{fig:Figure 3}f, although the spread is much narrower across $\lambda_x$ compared to the corresponding near-wall case (Figure \ref{fig:Figure 3}c). This demonstrates that compliance-induced effects become weaker as modes move further away from the wall. The compliance-induced effects peak for $\lambda_x/H = 3$ for both $H = 0.1$ and $H = 0.25$, further demonstrating the strong scaling with coating thickness. Advecting structures at $c = 0.6U_\infty$ with similar wavelengths were observed by experimental studies of \cite{wang2020interaction}. The original rigid wall low rank region remains undisturbed for both of the cases. Finally, moving on to softer surfaces with $E = 0.01$ in Figure \ref{fig:Figure 3}h, the vertical bands disappear due to those wavelengths being unstable. Moreover, the dominance of the first resolvent mode reduces for certain wavelength combinations centered around larger scales. This is in contrast to Figure \ref{fig:Figure 3}d, where the near-wall cycle remained nearly unaffected. While the dominance of the first resolvent mode is reduced, the region still remains relatively low rank with $R \approx 0.7$. Thus, in all the above cases across different values of $E$ and $H$, the resolvent operator remains low rank for $(\lambda_x, \lambda_z)$ pertaining to known turbulent motions. 

\begin{figure}
\centering
      \includegraphics[width=1\linewidth]{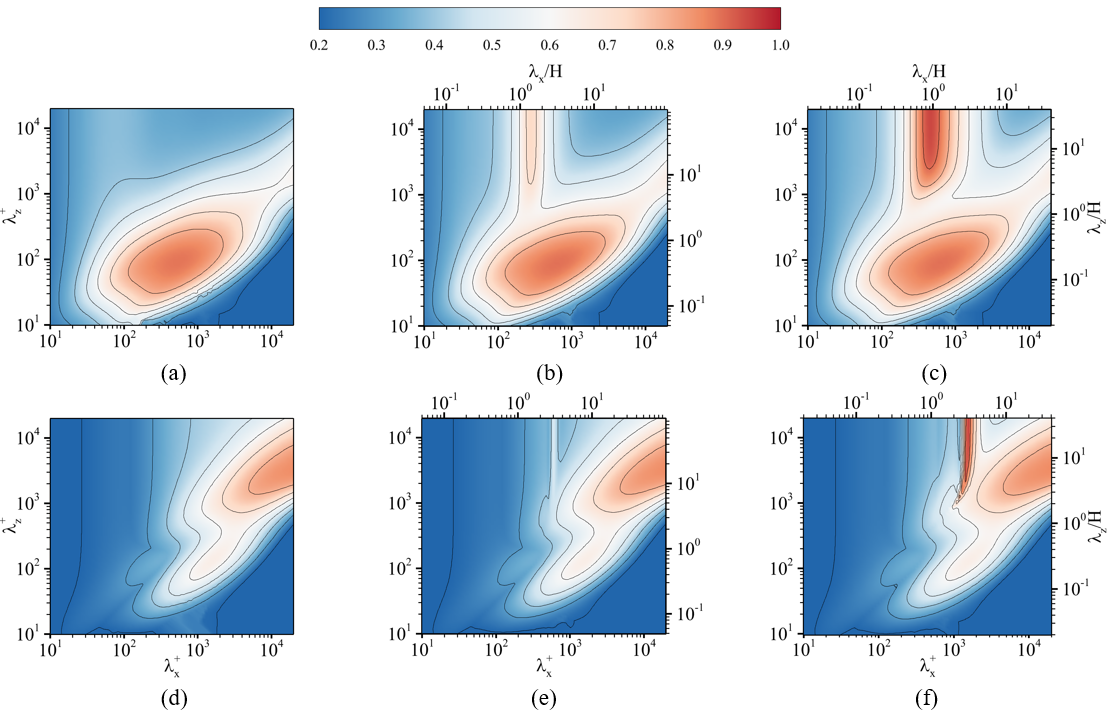}
    \caption{Fraction of energy captured by the first resolvent mode $\sigma^2_1/\sum_{i}\sigma_i^2$ across different wavelength combinations $(\lambda^+_x, \lambda^+_z)$ using eddy viscosity for at $c^+ = 10$ (a-c) and $c^+ = 16$(d-f).  The cases presented are \emph{(a,d)} Rigid, \emph{(b,e)} $E = 0.5, H = 0.1$ and \emph{(c,f)} $E = 0.5, H = 0.25$. The solid viscosity and density are maintained constant as $\mu_s = 0.005$ and $\rho_s = 1$. } 
\label{fig:Figure 4}
\end{figure}

Figure \ref{fig:Figure 4} shows the low rank spectrum using an eddy viscosity resolvent. Figures \ref{fig:Figure 4}(a-c) are generated using similar parameters as Figures \ref{fig:Figure 3}(a-c), while Figures \ref{fig:Figure 4}(d-f) are identical to Figures \ref{fig:Figure 3}(e-g). The rigid wall cases for $c^+ = 10$ and $c^+ = 16$ show distinct low rank regions corresponding to the near-wall cycle and large scale modes respectively, which are identical to what was observed with the standard resolvent operator. However, the low rank regions with eddy resolvent are relatively smaller and more distinct, with $R \approx 0.8$. As the wall is made compliant for $c^+ = 10$ (Figure \ref{fig:Figure 4}(b,c)), a separate low rank branch appears which corresponds to the fluid-structure coupling, and it occurs for a constant $\lambda_x^+$ and $\lambda_z^+>1000$. The dominant wavelength scales with $H$, occurring for values of $1<\lambda_x/H<2$, and the its dominance becomes stronger with increasing $H$. The near-wall region remains unchanged. This is very similar to what was observed in Figure \ref{fig:Figure 3}.  For $c^+ = 16$, the observations remain similar. The fluid-structure coupling band is observed for $\lambda_x/H = 3$, while the large scale modes remain unchanged. Overall, it can be concluded that the dominance of the leading singular value remains qualitatively similar spectrally, whether an eddy viscosity model is used or not. However, the resolvent operator becomes slightly higher rank across all these scales when eddy viscosity is included.   

\begin{figure}
	\centering
    	\begin{subfigure}[t]{0.49\textwidth}
		\includegraphics[width=\textwidth]{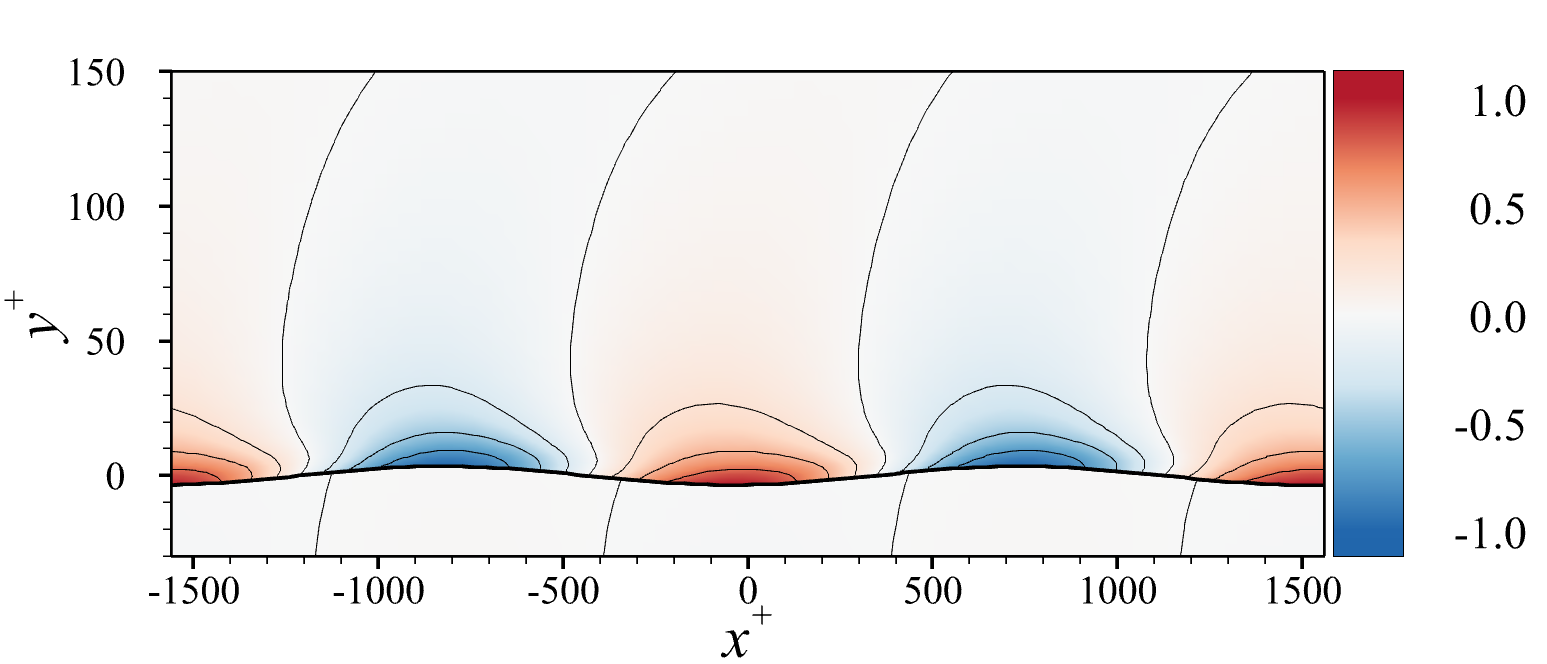}
		\caption{}
		\label{fig:5a}
	\end{subfigure}
	\hfill
	\begin{subfigure}[t]{0.49\textwidth}
		\includegraphics[width=\textwidth]{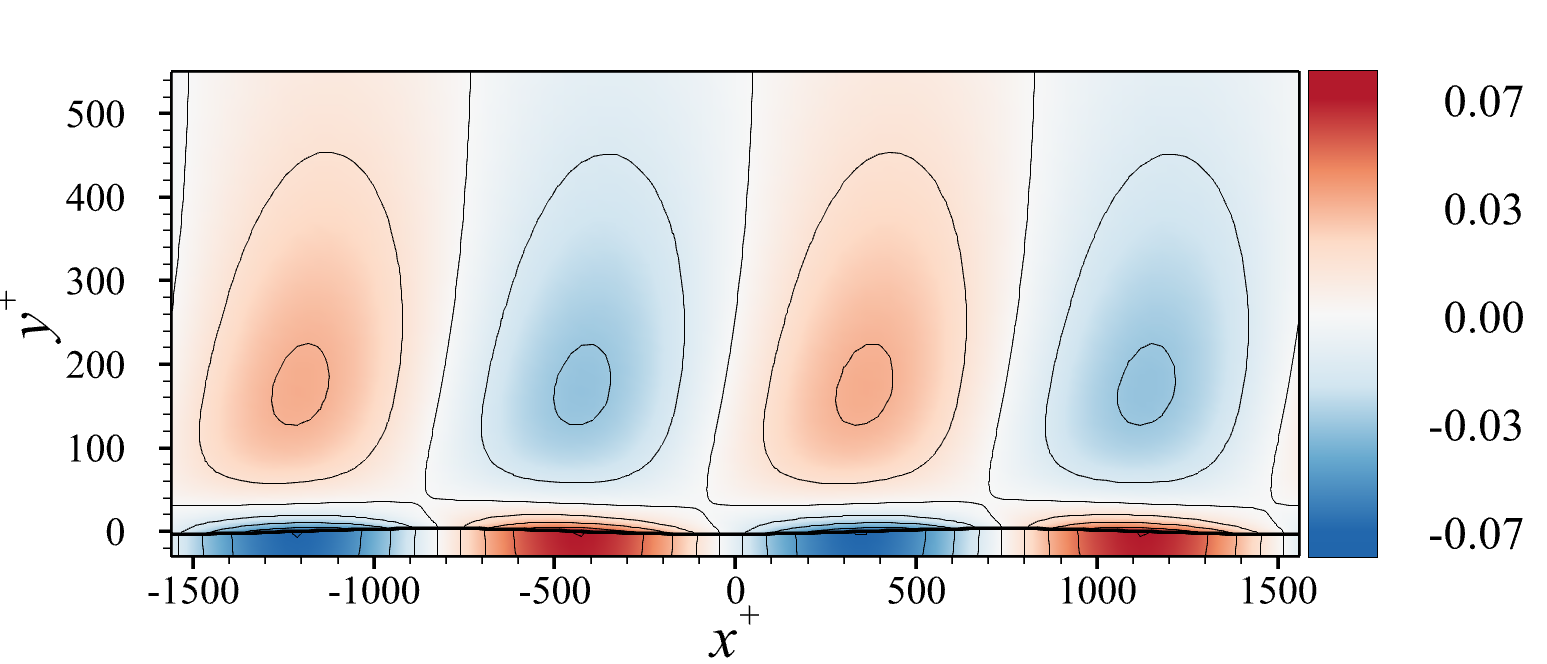}
		\caption{}
		\label{fig:5b} 
	\end{subfigure}	
\caption{Response mode structures in the $x-y$ plane for compliant surface $E = 0.5$, $H = 0.25$ and $(\lambda_x^+,\lambda_z^+,c^+) = (1500, 10000, 16)$, marked in Figure \ref{fig:Figure 4}f. \emph{(a)} $u$ response, \emph{(b)} $v$ response.} 
\label{fig:Figure 5}
\end{figure}

Figure \ref{fig:Figure 5} shows the response structures for the resolvent mode from one of the vertical bands $(\lambda_x^+, \lambda_z^+,c^+) = (1500,10000,16)$ in Figure \ref{fig:Figure 4}f in the streamwise wall-normal plane. These high-gain resolvent modes are driven by the fluid-structure coupling and are two-dimensional in nature, as can be observed by their high $\lambda_z$ cutoff in Figures \ref{fig:Figure 3} and \ref{fig:Figure 4}. Figure \ref{fig:5a} shows the $u$ response of the fluid and solid in deformed coordinates. It can be seen that the streamwise velocity in the fluid domain peaks at the surface, but the solid velocity is negligible compared to the fluid. The strong $u$ response is due to the linearized kinematic condition [\ref{eq:15}], where streamwise velocity of the fluid near the surface is driven by the coupling between strong mean velocity gradient at $y = 0$ and wall-normal deflection of the compliant surface. Results by \cite{wang2020interaction} and \cite{lu2025analysis} show a significant reduction in the mean velocity gradient near the wall compared to a rigid wall case, which in turn could reduce predicted streamwise velocities near the interface. Figure \ref{fig:5b} shows the $v$ response. Compared to $u$, the $v$ response in the fluid extends much further from the wall, peaking near $y^+ = 160$. Compared to the fluid, the solid velocity is much higher, and organized in phase opposition to the fluid response. For modes with $c^+  = 10$ (Figure \ref{fig:Figure 4}a, \ref{fig:Figure 4}b), the response shapes remain qualitatively similar (not shown), but the $v$ peak is found to be at $y^+ = 50$. The $v$ peak being closer to the wall for $c^+ = 10$ may lead to stronger pressure fluctuations at the surface \citep{luhar2014structure}, which would in turn lead to the relatively stronger coupling observed in cases with $c^+ = 10$ compared to $c^+ = 16$.

\subsection{Interaction with near-wall cycle}

In this section, we shall focus on the effect of viscoelastic compliant surface interaction with modes that represent the near-wall cycle (NW) in turbulent boundary layers. Within the framework of resolvent analysis, control approaches for the near-wall cycle have been undertaken \citep{luhar2015framework, chavarin2020resolvent, chavarin2021resolvent, jafari2023frequency} to assess the effects of spring-backed compliant walls, riblets, and wall permeability. As discussed in the previous section, the modes characterized by the triplet $(\lambda_x^+, \lambda_z^+, c^+) = (1000, 100, 10)$ serve as a good proxy for the near-wall cycle. Since an isotropic viscoelastic surface is governed by four independent parameters $(E, H, \rho_s, \mu_s)$, we would only vary two parameters simultaneously $(E, H)$ for the current discussion. The effects of $\mu_s$ and $\rho_s$ are also important and will be discussed in subsequent sections, and they will not affect the qualitative conclusions obtained in this section. The density ratio will be maintained as $\rho_s = 1$, which is ideal for the interaction of viscoelastic surfaces with low speed hydrodynamic flows. For the damping term, we maintain a constant loss tangent $\phi$ instead of using a constant viscosity $\mu_s$. The loss tangent is defined as $\tan (\phi) = \omega \mu_s/E$ for a viscoelastic material following the Kelvin-Voigt model. Since we are looking at a single mode with constant circular frequency $(\omega^+ = 2\pi c^+/\lambda^+_x)$, the constant loss model ensures that the variation of $E$ does not produce impractical values of viscoelastic damping compared to its storage modulus. For practical moderate viscosity elastomers and rubbers, the loss tangent of $\tan(\phi) = 0.1$ serves as a good estimate and is used for the present study. 

In order to assess the effectiveness of the compliant surface for modulation of the given mode, two metrics will be used. The ratio of resolvent gains between compliant and rigid surfaces $\sigma_{\textbf{k}c}/\sigma_{\textbf{k}0}$ provides a direct measure to assess the amplification or attenuation of the overall mode kinetic energy. To assess the effect of the compliant surface on turbulent drag, the Renard-Deck decomposition \citep{renard2016theoretical} of turbulent skin friction is used. The decomposition uses the kinetic energy budget to express the $C_f$ in terms of energy dissipation through mean velocity gradients, turbulent kinetic energy production, and streamwise development of the velocity profile. The integrated turbulent kinetic energy (TKE) production term from the budget equation is used to express the drag associated with the mode. In the resolvent framework, this is given by \citep{luhar2015framework, renard2016theoretical}
\begin{align}
    P_{\textbf{k}} = \sigma_\textbf{k}^2\int_{y = 0}^{\infty} \text{Re}(-u_\textbf{k}^*v_\textbf{k})\bar{\rho}\frac{d\bar{U}}{dy} dy
     \label{eq:29}
\end{align}
Here $u_\textbf{k}, v_{\textbf{k}}$ are the normalized velocity response modes. The terms TKE production and Reynolds stress will be used interchangeably in the subsequent sections. The total drag can be quantified by integrating the Reynolds shear stress component for each mode across $\textbf{k}$. It can be seen that $P_\text{k}$ depends strongly on both $\sigma_{\text{k}}$ and the inner product of $u$ and $v$. Since the integrand contains the mean velocity gradient term, changes in mode reorganization near the wall would strongly affect the TKE production. 

\begin{figure}
	\centering
    	\begin{subfigure}[t]{0.31\textwidth}
		\includegraphics[width=\textwidth]{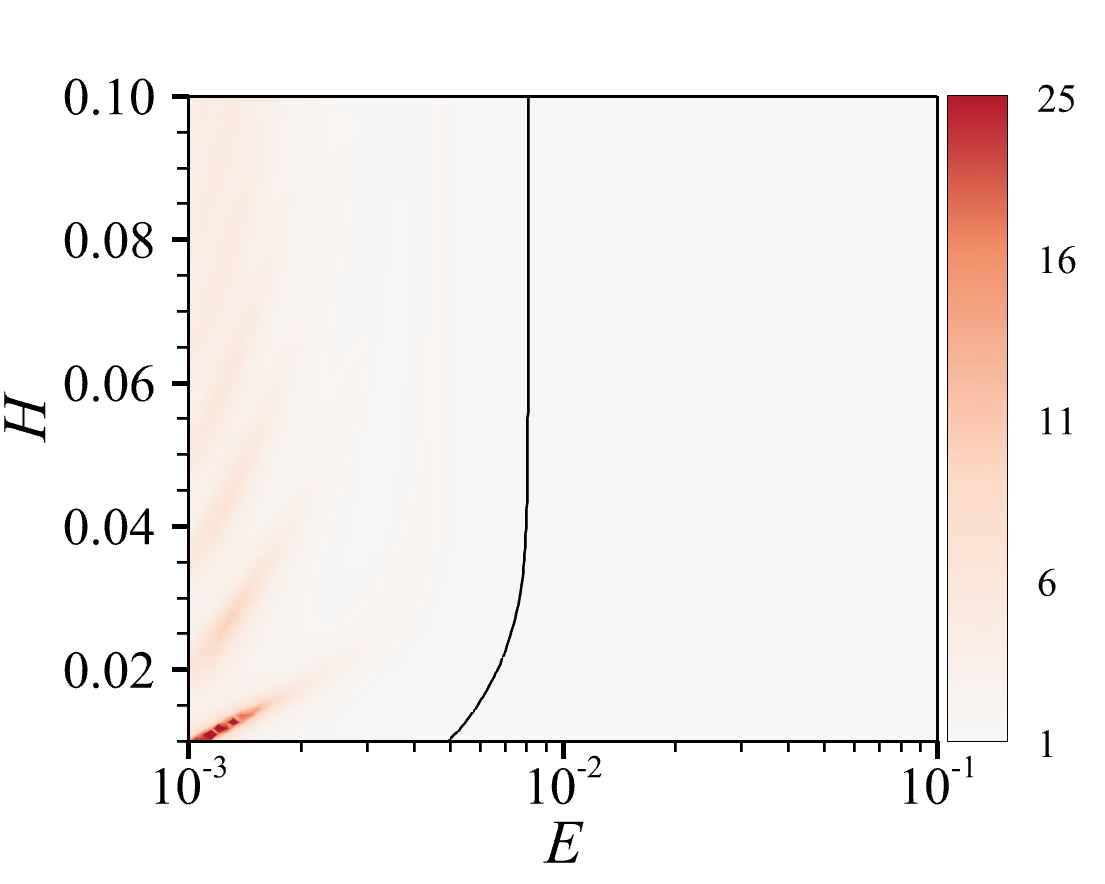}
		\caption{}
		\label{fig:6a}
	\end{subfigure}
	\hfill
	\begin{subfigure}[t]{0.31\textwidth}
		\includegraphics[width=\textwidth]{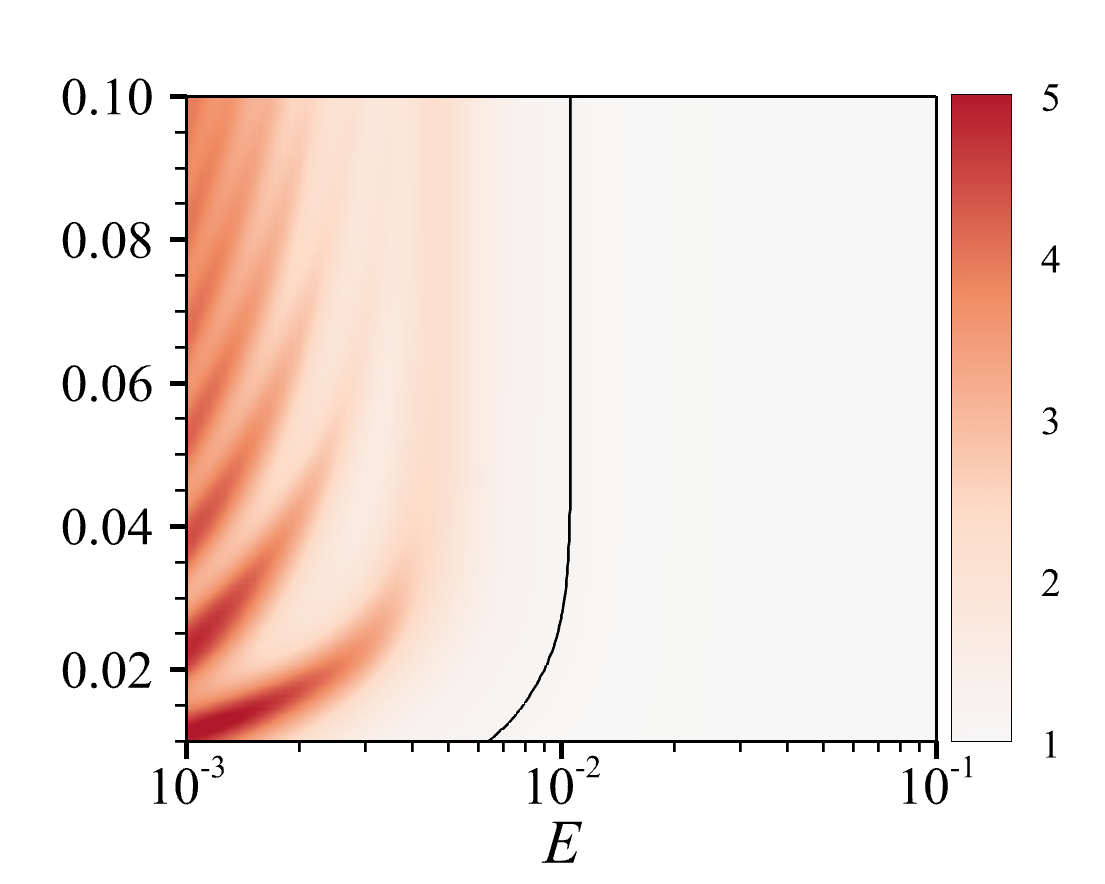}
		\caption{}
		\label{fig:6b} 
	\end{subfigure}	
        \hfill
        \begin{subfigure}[t]{0.31\textwidth}
				\includegraphics[width=\textwidth]{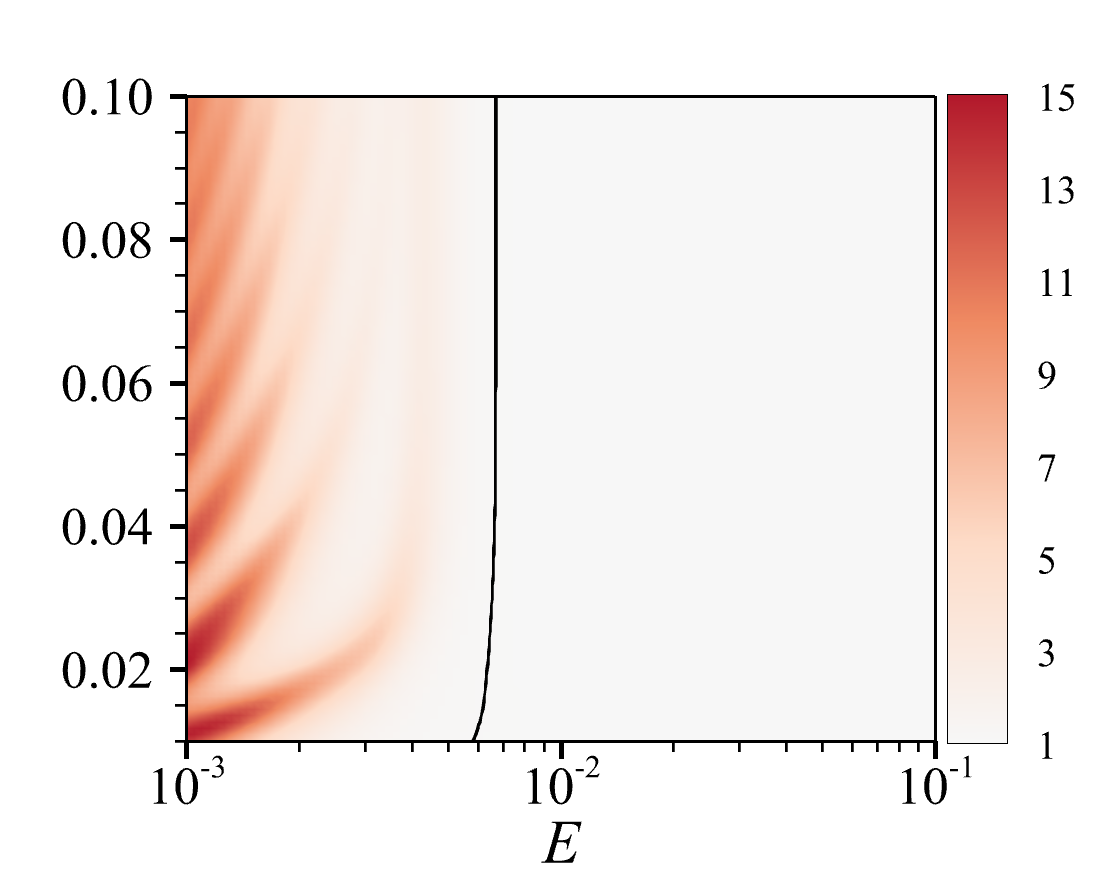}
		\caption{}
		\label{fig:6c} 
	\end{subfigure}	
        \hfill
\caption{Contour plots spanning $E$ and $H$ showing the effect of resolvent gain of NW modes due to interaction with viscoelastic surface \emph{(a)} Compliant to rigid wall singular values $\sigma_{\textbf{k}c}/\sigma_{\textbf{k}0}$ for standard resolvent, \emph{(b)}  Compliant to rigid wall singular values $\sigma_{\textbf{k}c}/\sigma_{\textbf{k}0}$ for eddy resolvent, and \emph{(c)} Compliant to rigid wall production $P_{\textbf{k}c}/P_{\textbf{k}0}$ for eddy resolvent. The other parameters $\rho_s = 1, \tan \phi = 0.1$ remain constant. The solid contour lines denote the locus of points where the contour value is unity.} 
\label{fig:Figure 6}
\end{figure}    

Figure \ref{fig:6a} shows the resolvent gain ratio $\sigma_{\textbf{k}c}/\sigma_{\textbf{k}0}$ across a range of normalized storage modulus $E$ and coating thickness $H$ using the standard resolvent. All the cases investigated are linearly stable across all wavelengths. It can be seen that viscoelastic coatings with $E>0.01$ do not couple with the NW mode, resulting in an amplification of unity. Strong interaction is found to occur in discrete bands in the $E -H$ space, starting below the threshold of $E = 0.01$. This threshold increases with solid thickness $H$ until $H = 0.04$, and becomes invariant with solid thickness for $H > 0.04$. The interactions occur along discrete bands which become nearly invariant with $H$ for large $H$. Even though strong interactions are observed for multiple coating combinations, all of them result in increased amplification of the NW mode compared to the baseline. As $E$ or $H$ is reduced along a given band, the energy amplification increases. For thicknesses beyond $H = 0.1$, the threshold $E$ remains invariant, and the interactions gradually weaken with increasing $H$. Figure \ref{fig:6b} shows the eddy resolvent gain ratio $\sigma_{\textbf{k}c}/\sigma_{\textbf{k}0}$ for the NW mode. Similar to what was observed in Figure \ref{fig:6a}, the energy of the NW mode with eddy viscosity included is amplified by fluid-structure coupling; however, the magnitude of amplification is much weaker. The qualitative topology of the interaction remains similar to Figure \ref{fig:6a}. Finally, the ratio of production $P_{\textbf{k}c}/P_{\textbf{k}0}$ for eddy resolvent is shown in Figure \ref{fig:6c}. Similar to Figure \ref{fig:6b}, all combinations of $E$ and $H$ exhibit an increase in turbulent production, such that combinations with larger $P_{\text{k}c}$ coincide with regions of large  $\sigma_{\textbf{k}c}$. This is due to the direct relation between production and resolvent gain [\ref{eq:29}], where the large increase in resolvent gain overpowers any favorable reorganization of the response $u$ and $v$ modes. The contours generated using the standard resolvent are also qualitatively similar and hence are not shown. Overall, it can be concluded that viscoelastic coatings do not produce a reduction in resolvent gain or Reynolds stress when interacting with near-wall streamwise vortices.

\subsection{Interaction with large scale motions}

\begin{figure}
\centering
      \includegraphics[width=1\linewidth]{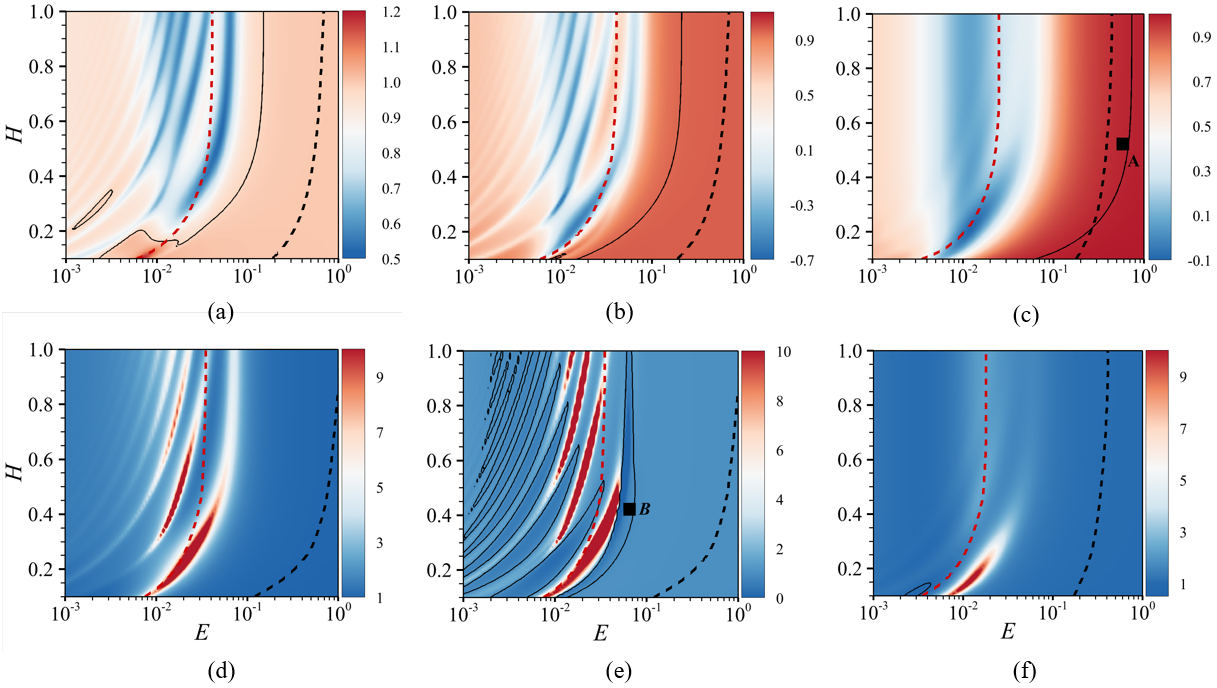}
    \caption{Contour plots showing the ratios of \emph{(a,d)} Compliant to rigid wall singular values $\sigma_{\textbf{k}c}/\sigma_{\textbf{k}0}$ for $\tan \phi = 0.1$, \emph{(b,e)} Compliant to rigid wall TKE production $P_{\textbf{k}c}/P_{\textbf{k}0}$ for $\tan \phi = 0.1$, and \emph{(c,f)} Compliant to rigid wall TKE production $P_{\textbf{k}c}/P_{\textbf{k}0}$ for $\tan \phi = 0.5$. Figures \emph{(a-c)} were obtained using traditional resolvent analysis, whereas \emph{(d-f}) were obtained using the eddy viscosity profile. The dashed black line denotes the critical $E$ for onset of traveling wave flutter instabilities for the respective cases. The solid black line denotes values of $E$ where the contour value is unity. The red dashed line denotes cases where the coupled linearized system is unstable for $(\lambda_x,\lambda_z) = (4,1)$.} 
\label{fig:Figure 7}
\end{figure}

In this section, a similar approach as discussed in the previous section shall be used to discuss the interaction between the viscoelastic compliant surface and large scale motions in the boundary layer. Large scale motions in the boundary layer typically have spanwise extents of $\delta$ and streamwise lengths within 4 $\delta$ -- 10 $\delta$, convecting at speeds around $0.6 U_\infty - 0.7 U_\infty$. For this analysis, we use the triplet $(\lambda_x, \lambda_z, c) = (4, 1, 0.6)$ to represent the large scale mode. Figure \ref{fig:Figure 7} shows the resolvent gain ratio and turbulent production ratio contours for materials of different $E$ and $H$, using both standard (Figures \ref{fig:Figure 7}(a-c)) and eddy viscosity based (Figures \ref{fig:Figure 7}(d-f)) resolvent analysis. The loss tangent and density ratio are maintained to be $\tan \phi = 0.1$ and $\rho_s = 1$ respectively. In this case, for certain combinations of solid parameters, the coupled system is linearly unstable at ($\lambda_x,\lambda_z$) = (4,1), and is demarcated by the red dashed lines in Figure \ref{fig:Figure 7}. However, the real part of the unstable eigenvalue ($\omega_r = 0.6$) and the mode frequency at which the resolvent operator is forced ($\omega = 0.92$) are significantly spaced apart, hence the effect of the unstable mode is negligible at the frequency of interest.

The ratio of compliant to rigid resolvent gain (Figure \ref{fig:Figure 7}a) shows that the majority of materials in the range $0.005<E<0.09$ and $0.2<H<1$ result in significant attenuation of resolvent gain by at most 50\%. The solid line marks regions where the contour value is unity, thus denoting the boundary beyond which the viscoelastic surface becomes unresponsive to these scales. Strong interactions occur along discrete bands across $E$ and $H$, which are separated by regions of weaker interactions. Along the bands, with increasing $H$, the interactions are inclined towards lower $E$. The contours of TKE production are shown in Figure \ref{fig:Figure 7}b. For $E<0.1$ and $H>0.2$, a decrease in $P_{\textbf{k}c}$ is observed for all combinations of $E$ and $H$. Within $0.006<E<0.08$, certain bands exist in which $P_{\textbf{k}c}$ shows a reversal of sign, which can occur when $u$ and $v$ modes become positively correlated near the wall. These bands approximately coincide with the regions showing strong reduction in $\sigma_{\text{k}c}$ in Figure \ref{fig:Figure 7}a. The solid black line shows that for $E>0.2$, the compliant wall is relatively ineffective at reducing Reynolds stress.

The dashed black solid lines plotted in Figure \ref{fig:Figure 7}(a,b) represent the neutral curves for the onset of convective instabilities in the viscoelastic layer for spanwise uniform ($k_z = 0$) modes. This curve was obtained from a separate linear stability analysis using $\rho_s = 1$ and $\mu_s = 0.005$, representing weakly damped viscoelastic layers. 
%The green curve denotes the neutral curve for optimally damped viscoelastic layers for a given combination of $E$ and $H$, and is discussed further in Figure \ref{fig:Figure 8}. 
The choice of modes with $k_z = 0$ is made since two-dimensional disturbances are amplified the strongest \citep{yeo2001turbulent}. The onset of convective instabilities on viscoelastic layers (commonly referred to as traveling wave flutter) has been studied extensively \citep{gad1986response, yeo1990hydrodynamic, yeo2001turbulent, pfister2022global, greidanus2022response, esteghamatian2022spatiotemporal}. These waves travel at speeds $c \approx C_t$ on the compliant surface, and have commonly been reported to increase drag \citep{greidanus2022response, esteghamatian2022spatiotemporal,wang2020interaction,zhang2017deformation}. It can be observed from Figure \ref{fig:Figure 7}(a,b) that the critical modulus at which fluid-structure instabilities onset is an order of magnitude higher than the values of $E$ where strong favorable interactions ($E<0.1$) are expected with the large scale modes. Thus, the onset of fluid-structure instabilities precedes the potential favorable interactions, making the drag reduction prospect of weakly damped viscoelastic materials practically unrealizable.

One of the ways to delay the onset of flutter is to increase the damping or viscosity of the solid. Figure \ref{fig:Figure 7}c shows the TKE production $P_{\textbf{k}}$ ratio for a material with $\tan{\phi} = 0.5$, i.e., a more viscous coating compared to Figure \ref{fig:Figure 7}b. Although increasing viscosity results in a much weaker interaction compared to the previous case, the reduction in $P_\textbf{k}$ still remains significant. The extended regions where strongly negative Reynolds shear stress were observed in Figure \ref{fig:Figure 7}b have shrunk and become more localized around $H = 0.2$ and $E = 0.02$. However, due to the effect of larger viscosity, the threshold for $P_{\text{k}c}/P_{\text{k}0}<1$ has extended beyond the flutter threshold to larger values of $E \approx 0.8$. The flutter threshold in this case was obtained using $\mu_s = 0.05$. It can be seen that the threshold for flutter has shifted towards lower $E$, i.e., the onset of flutter is delayed due to stronger damping. Interestingly, for values of $E$ larger than the critical values at the flutter boundary, a reduction of $P_\text{k}$ by up to $4\%$ can still be observed.  

Figure \ref{fig:Figure 7}d shows the resolvent gain ratio obtained for viscoelastic coatings with $\tan{\phi} = 0.1$ using the eddy viscosity profile. In contrast to the results shown using traditional resolvent analysis in Figure \ref{fig:Figure 7}a, inclusion of the eddy viscosity results in an increase in resolvent gain ratio across the entire range of $E$ and $H$. Regions of strong amplification ($\sigma_{\text{k}c}/\sigma_{\text{k}0} > 10$) are observed along discrete bands between $0.006<E<0.1$. Figure \ref{fig:Figure 7}e shows the ratio of TKE production using the eddy viscosity profile. Since the TKE production ratio depends strongly on the resolvent gain ratio [\ref{eq:29}], regions of high $\sigma_{\text{k}c}$ and $P_{\text{k}c}$ overlap with each other. However, banded regions with reduced TKE production ($P_{\text{k}c}/P_{\text{k}0}<1$) still exist alongside the amplified regions, as shown within the enclosed by the solid lines. These regions are much smaller in extent as compared to the standard resolvent case (Figure \ref{fig:Figure 7}b). 

%The difference in amplification behavior between standard and eddy resolvent cases can be explained by how the mode energy is distributed in the wall-normal direction. Since the wall pressure is the dominant forcing term on the solid surface, \cite{luhar2014structure} showed that the the wall pressure signature 

While response mode shapes with standard resolvent exhibit strong narrow peaks localized near the critical layer, the response mode shapes of VLSMs with eddy resolvent are spread over a larger wall-normal extent, i.e., they have a stronger near-wall footprint. Moreover, inclusion of eddy viscosity reduces the anisotropy between the different velocity components, resulting in a stronger $v$ response compared to standard resolvent. Both these factors can contribute to a stronger coupling between the VLSM mode and the viscoelastic surface when eddy viscosity is included, as discussed in the Appendix. The black dashed lines in Figure \ref{fig:Figure 7}d and \ref{fig:Figure 7}e denotes the critical $E$ for flutter onset, obtained using linear stability analysis using an eddy viscosity profile. Compared to the cases without eddy viscosity in Figure \ref{fig:Figure 7}(a,b), the critical $E$ is similar for lower $H$, but increases rapidly between $0.1<H<0.6$. Thus, inclusion of eddy viscosity with lower $\mu_s$ results in an earlier onset to flutter (i.e., at lower flow speeds or higher $E$). 

As the viscosity is increased to $\mu_s = 0.5$ as shown in Figure \ref{fig:Figure 7}f, the interactions become significantly weaker and diffuse, indicated by the disappearance of the strong banded structure. Moreover, the regions where a reduction in Reynolds shear stress was observed in Figure \ref{fig:Figure 7}e are also largely absent. Interestingly, the critical onset curve coincides strongly with that in Figure \ref{fig:Figure 7}c. This shows that addition of eddy viscosity does not have a strong effect on the onset of flutter for solids with higher $\mu_s$. The weak reduction in TKE production that was observed for $E>E_{\text{onset}}$ using standard resolvent was not observed when an eddy viscosity profile is used. 

\begin{figure}
	\centering
    	\begin{subfigure}[t]{0.31\textwidth}
		\includegraphics[width=\textwidth]{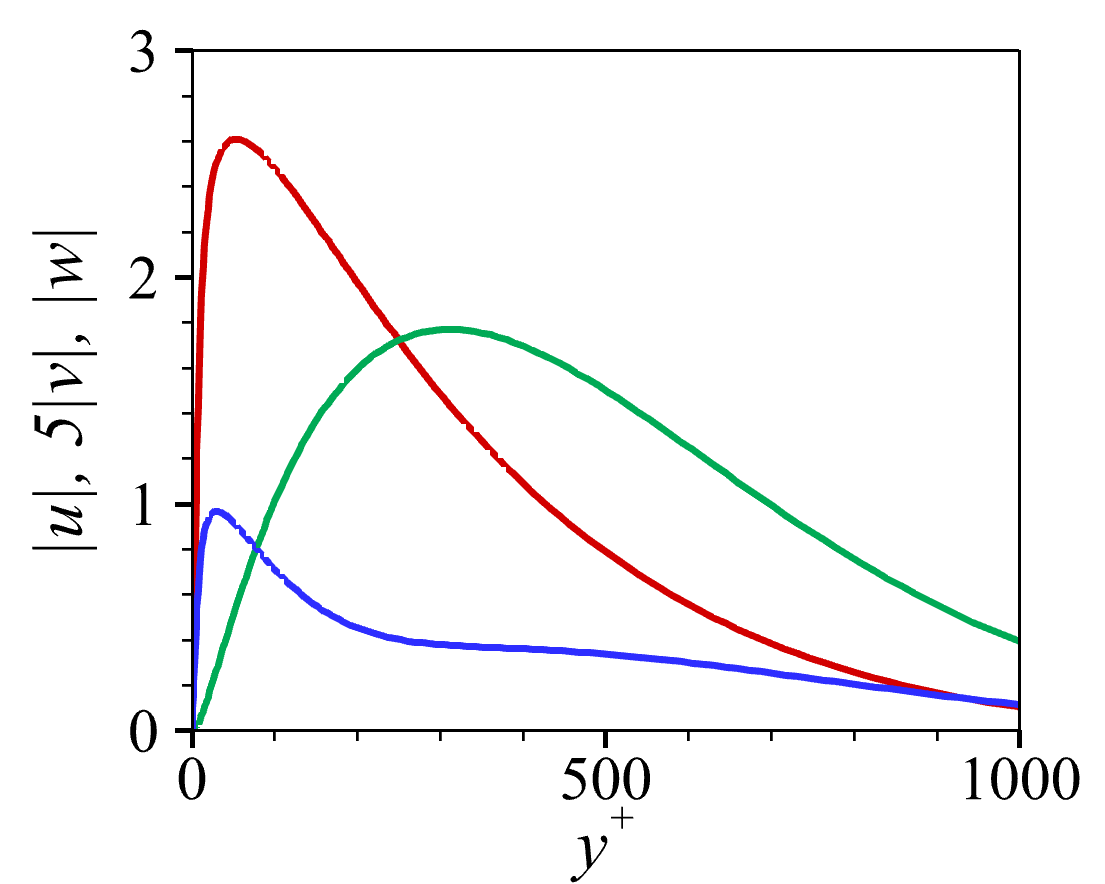}
		\caption{}
		\label{fig:8a}
	\end{subfigure}
	\hfill
	\begin{subfigure}[t]{0.31\textwidth}
		\includegraphics[width=\textwidth]{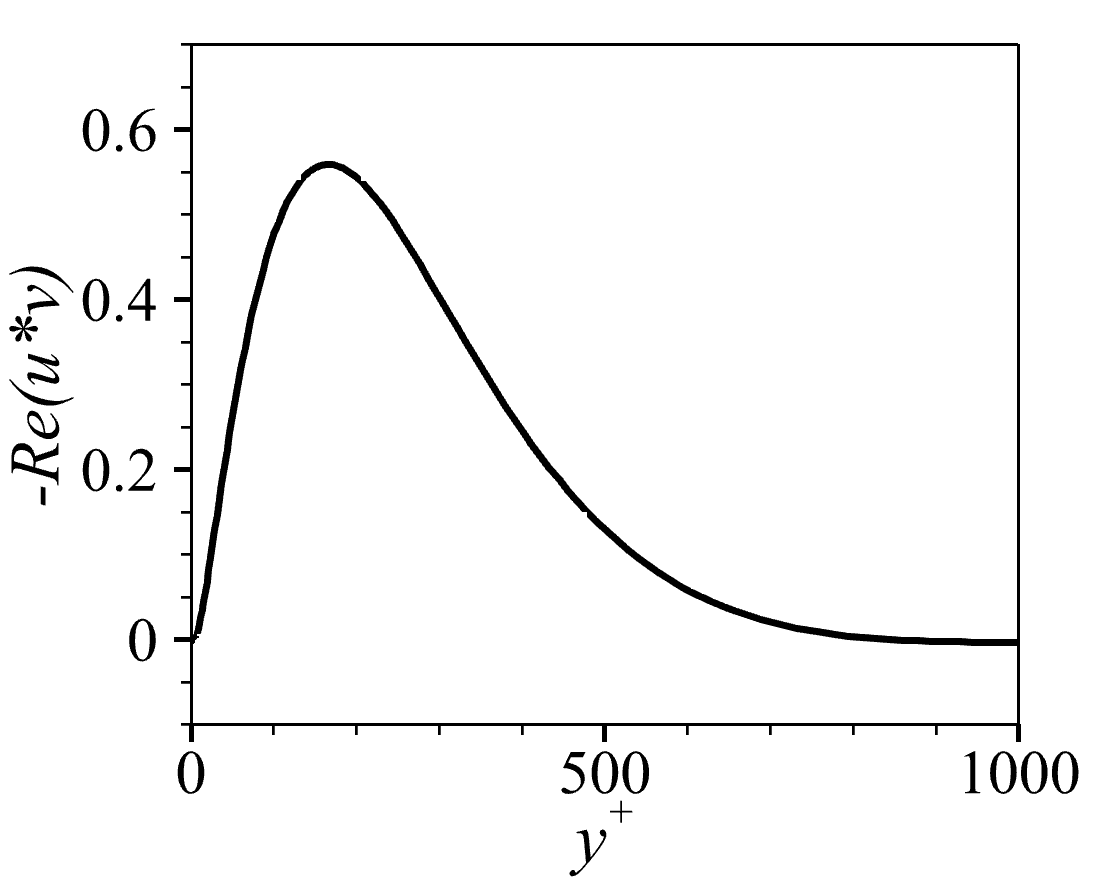}
		\caption{}
		\label{fig:8b} 
	\end{subfigure}	
        \hfill
        \begin{subfigure}[t]{0.31\textwidth}
		\includegraphics[width=\textwidth]{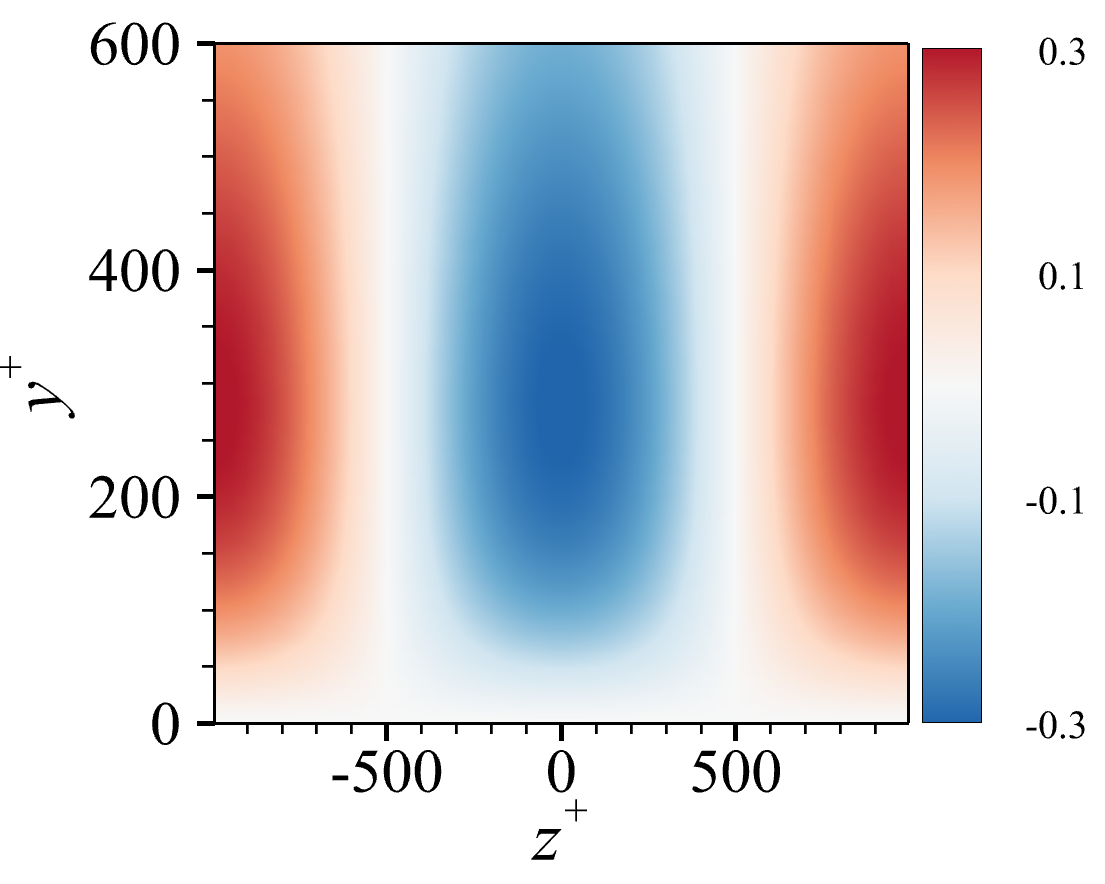}
		\caption{}
		\label{fig:8c} 
	\end{subfigure}	
        \hfill
        \begin{subfigure}[t]{0.31\textwidth}
		\includegraphics[width=\textwidth]{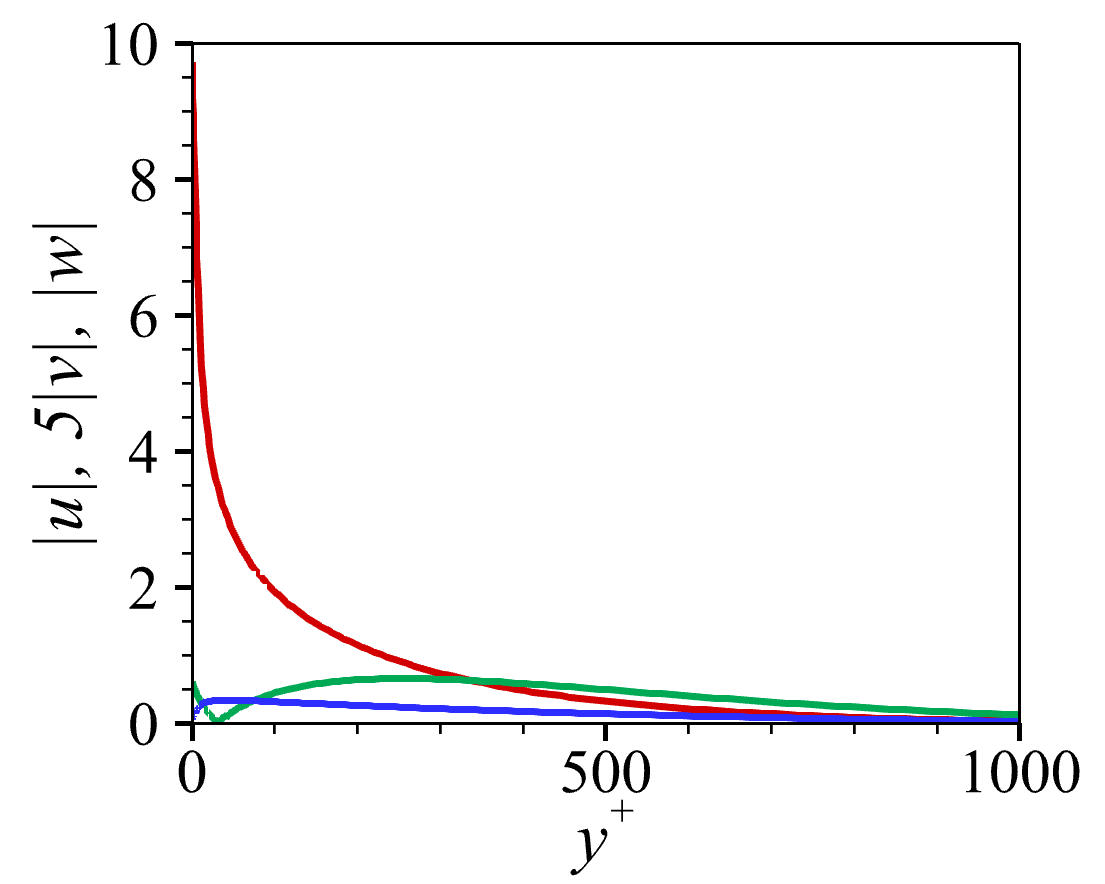}
		\caption{}
		\label{fig:8d}
	\end{subfigure}
	\hfill
	\begin{subfigure}[t]{0.31\textwidth}
		\includegraphics[width=\textwidth]{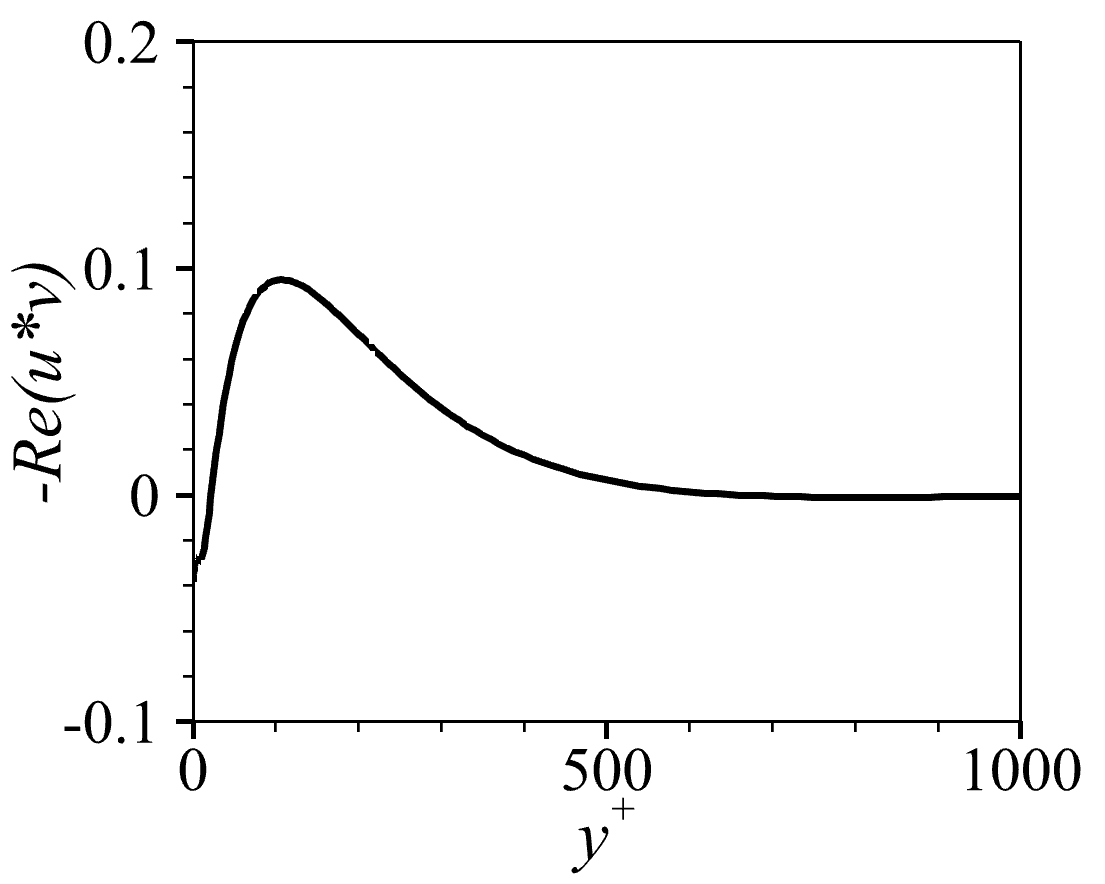}
		\caption{}
		\label{fig:8e} 
	\end{subfigure}	
        \hfill
        \begin{subfigure}[t]{0.31\textwidth}
		\includegraphics[width=\textwidth]{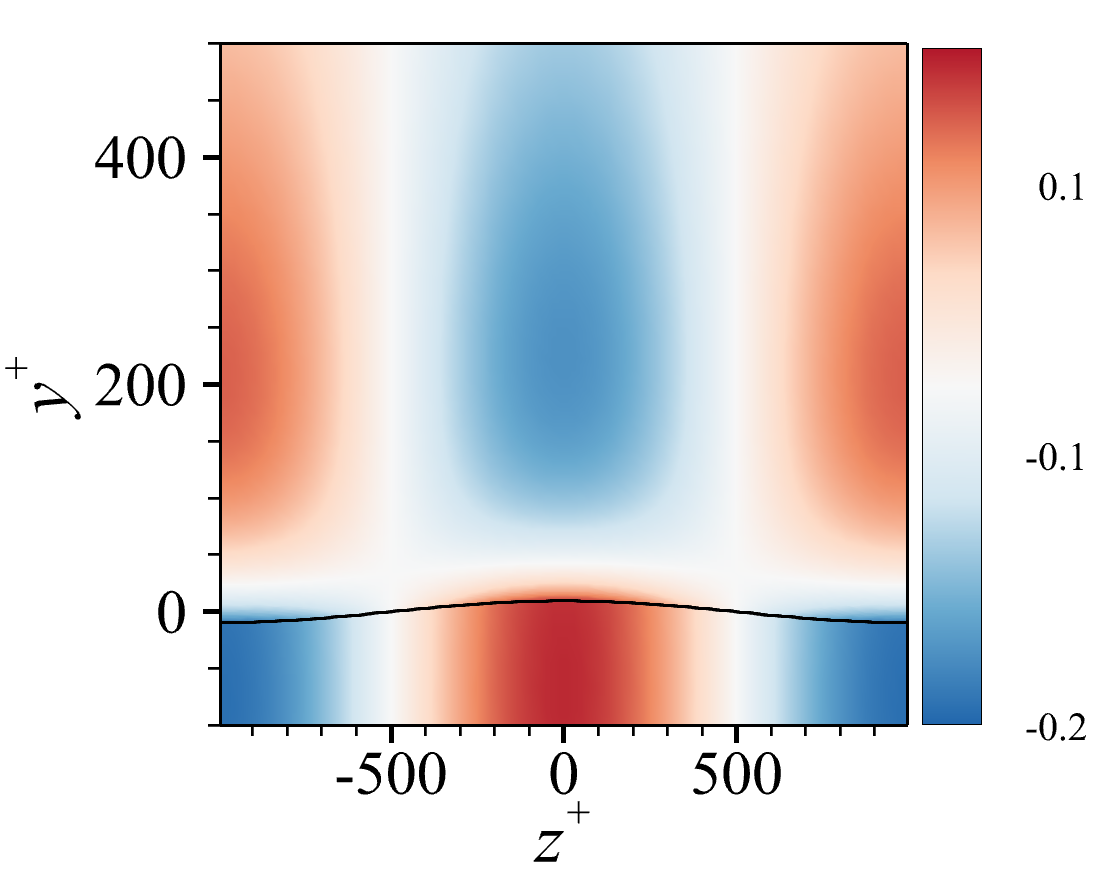}
		\caption{}
		\label{fig:8f} 
	\end{subfigure}	
        \hfill
\caption{wall-normal variation of response modes for rigid and viscoelastic walls for interaction with LSMs, using eddy viscosity resolvent. The viscoelastic case corresponds to marker $B$ in Figure \ref{fig:Figure 7}e. \emph{(a,d)} Velocity response amplitudes in the wall-normal direction with $u$ \colorline{red}, $v$ \colorline{green} and $w$ \colorline{blue}, \emph{(b,e)} Reynolds stress variation in the wall-normal direction, and \emph{(c,f)} Contours of wall-normal velocity $v$ in the $y-z$ plane. Panels \emph{(a-c)} correspond to rigid wall case, and \emph{(d-f)} correspond to viscoelastic case.} 
\label{fig:Figure 8}
\end{figure}

Figure \ref{fig:Figure 8} shows the wall-normal profiles of the velocity response for the VLSM mode for the rigid wall case and for a Reynolds stress reducing compliant wall case $B$ indicated in Figure \ref{fig:Figure 7}e. Figure \ref{fig:8a} shows that the $u$ and $w$ modes peak quite close to the wall, which is different from response modes obtained from standard resolvent, where modes peak near the critical layer. This is due to the enhanced turbulent transport caused by the wall-normal gradient of eddy viscosity \citep{symon2023use}, which transfers more energy towards the wall. The $v$ response is smaller and broader, and peaks far from the wall. The compliant wall velocity response is shown in Figure \ref{fig:8d}. The $u$ mode peaks at the interface due to the imposed linearized boundary condition given by [\ref{eq:15}], where the large mean velocity gradient at the wall creates a strong $u$ response at the wall. The $v$ response is broader and drops to zero at $y^+ = 30$, then increases towards the interface due to the wall compliance.

The Reynolds shear stress profiles are shown in Figures \ref{fig:8b} and \ref{fig:8e}. Near the interface below $y^+ = 40$, the compliant case shows negative Reynolds stress. Comparing between the rigid and compliant response, it is observed that the compliant case exhibits a much weaker peak which lies closer to the wall. Moreover, the Reynolds shear stress becomes negative at $y = 0$, which is due to non-zero streamwise velocity of the compliant surface. According to [\ref{eq:15}], if the streamwise solid velocity is constrained, the $u$ and $v$ components of the flow at $y =0$ become uncorrelated, resulting in zero Reynolds shear stress at the wall. The $v$ velocity contours are shown in the $y-z$ plane for both cases, in Figures \ref{fig:8c} and \ref{fig:8f}. For the compliant wall response, it can be seen that the viscoelastic surface responds exactly in phase-opposition to the fluid velocity, which can cause the observed weakening of the Reynolds shear stress of the mode. 

\subsection{Optimal Damping}

\begin{figure}
	\centering
    	\begin{subfigure}[t]{0.48\textwidth}
		\includegraphics[width=\textwidth]{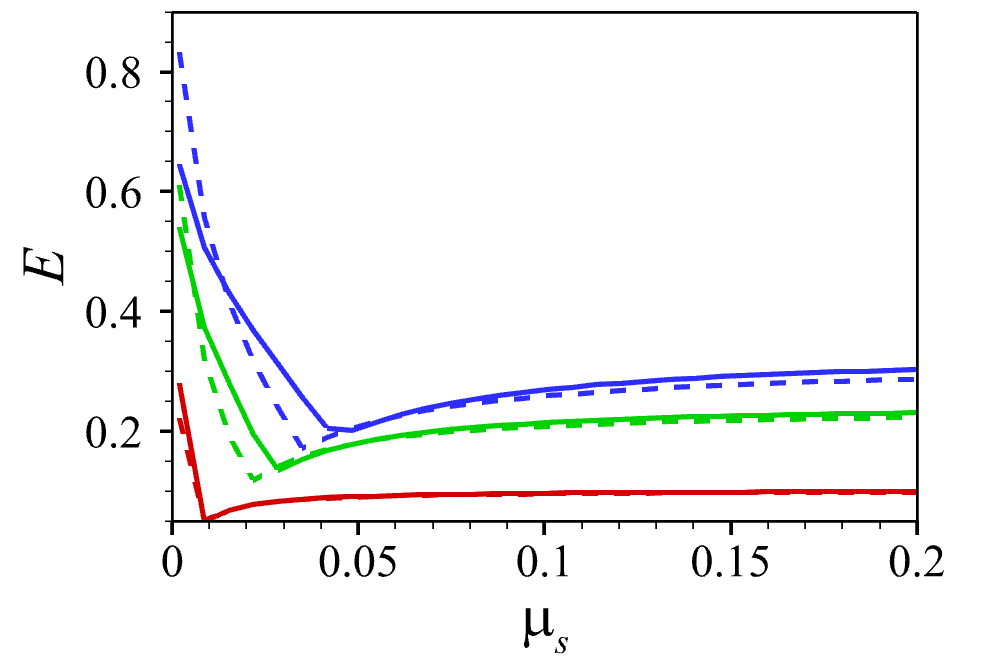}
		\caption{}
		\label{fig:9a}
	\end{subfigure}
	\hfill
	\begin{subfigure}[t]{0.48\textwidth}
		\includegraphics[width=\textwidth]{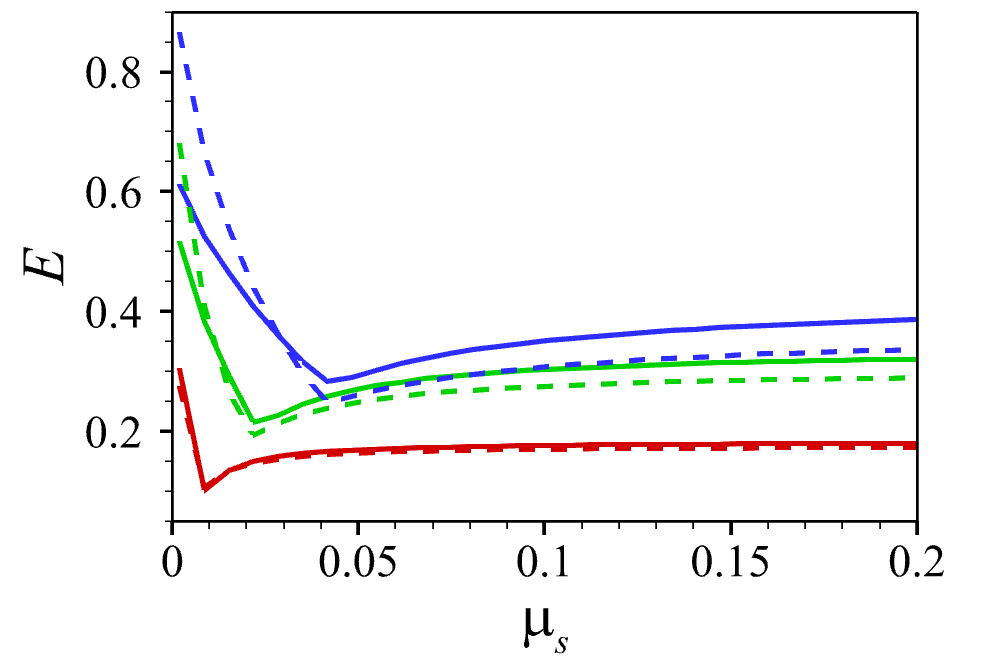}
		\caption{}
		\label{fig:9b} 
	\end{subfigure}	
        \hfill
    
\caption{Line plots showing $E_\textbf{critical}$ as a function of viscous damping $\mu_s$ for different coating thicknesses $H = 0.1$ \colorline{red}, $H = 0.3$ \colorline{green} and $H = 0.5$ \colorline{blue}, and different Reynolds numbers \emph{(a)}  $Re_{\tau} = 650$, and \emph{(b)}  $Re_{\tau} = 1990$. The solid lines are computed without an eddy viscosity model, whereas the dashed lines are computed using an eddy viscosity profile.}
\label{fig:Figure 9}
\end{figure}

Figure \ref{fig:Figure 7}c shows that materials with a high viscous damping spread the favorable interactions across a wide range of $E$, even though it weakens the interactions. The coupling still exhibits weak Reynolds stress reduction for materials with $E \approx 1$ (marked $A$ in Figure \ref{fig:Figure 7}c), which is higher than the critical values of $E$ at which linear stability analysis predicts the onset of flutter. It may therefore be desirable to delay the onset of flutter as much as possible by increasing the viscous damping $\mu_s$ to widen the margin between usable interactions and flutter onset. A larger margin would enable the use of solids with lower $E$, which can force stronger interactions with the turbulence. Figure \ref{fig:Figure 9} shows plots of $E_\textbf{critical}$ as a function of damping $\mu_s$ for coatings with different thicknesses $H$, with and without the inclusion of an eddy viscosity profile. For low values of damping, $E_{\textbf{critical}}$ is high but drops sharply, falling to a minimum, as expected with an increase in damping. The primary mode of instability in this range of $\mu_s$ is traveling wave flutter. With a further increase in damping, $E_\textbf{critical}$ increases weakly, and asymptotes to a steady value, which is slightly higher than the minimum value. In the high $\mu_s$ regime, the mode of instability is static divergence \citep{gad1984interaction, gad1986response, landahl1962stability, duncan1985dynamics, yeo2001turbulent}, which is absolute in nature. It can be seen that thicker coatings have much higher $E_\textbf{critical}$ across the entire range of $\mu_s$ i.e., thicker coatings will show onset to instability at lower flow speeds (or stiffer materials at a given flow speed). Moreover, the optimal damping at which a minimum $E_\textbf{critical}$ can be achieved increases with $H$. When an eddy viscosity profile is included, some notable differences can be observed. For $\mu_s<\mu_{s,\text{optimal}}$, the inclusion of eddy viscosity has a destabilizing effect, whereas for $\mu_s>\mu_{s,\text{optimal}}$, the eddy viscosity has a stabilizing effect on the onset of traveling wave flutter. At $\mu_s = \mu_{s,\text{optimal}}$, eddy viscosity has a very weak effect on the onset. Moreover, the effects of eddy viscosity decrease with decreasing thickness $H$, as can be seen for the curves corresponding to $H = 0.1$, where the curves from both formulations collapse.

Figures \ref{fig:9a} and \ref{fig:9b} show this for $Re_\tau$ of 650 and 1990, respectively. While both cases showcase similar qualitative trends, it can be seen that increasing $Re_{\tau}$ results in an earlier onset of instability (higher $E_\textbf{critical}$). The difference between the minimum $E_\text{critical}$ and the asymptotic value is slightly smaller than 0.1. From Figure \ref{fig:Figure 7}, the values of $E$ where strong favorable interactions may occur are at least ten times smaller than $E_{\text{onset}}$. This means that $\mu_s$ cannot be optimized enough for the favorable interactions to precede flutter. For higher values of $\mu_s$, it might still be possible to delay flutter slightly and utilize the possible sub-optimal interactions occurring in the stable regime. However, FISI exhibit significant transient growth \citep{pfister2022global,tsigklifis2017interaction} even in the stable regime, which may render these marginal interactions untenable. These interactions will be explored in the following sections.

\subsection{Effect of solid density} 

\begin{figure}
	\centering
    	\begin{subfigure}[t]{0.48\textwidth}
		\includegraphics[width=\textwidth]{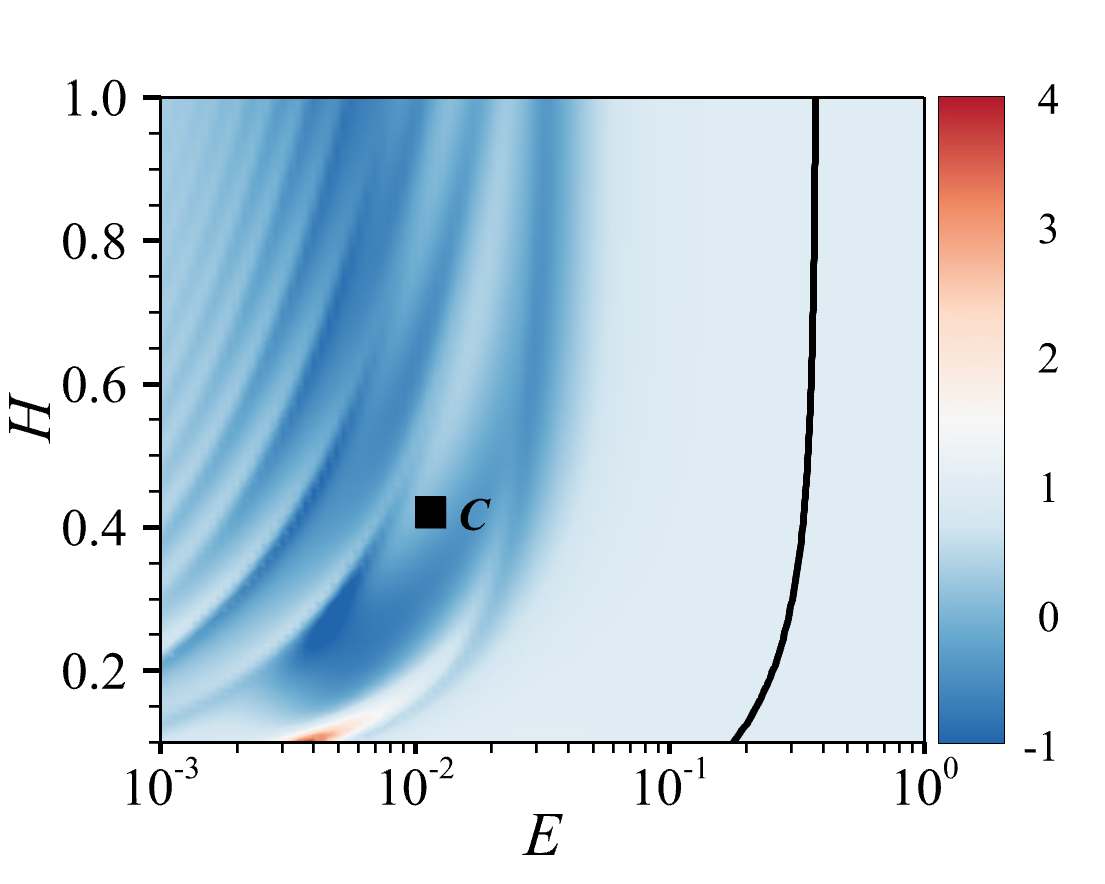}
		\caption{}
		\label{fig:10a}
	\end{subfigure}
	\hfill
	\begin{subfigure}[t]{0.48\textwidth}
		\includegraphics[width=\textwidth]{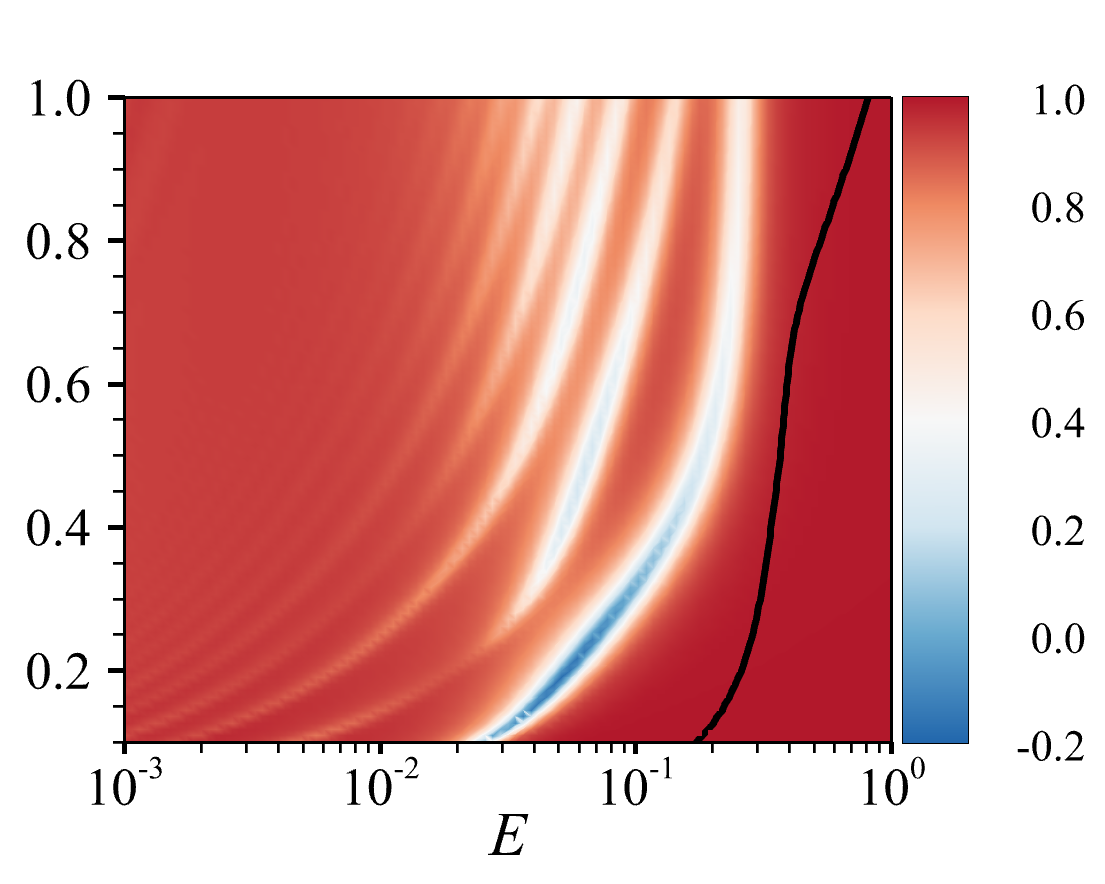}
		\caption{}
		\label{fig:10b} 
	\end{subfigure}	
        \hfill
\caption{Contour plots showing the ratios of compliant to rigid wall TKE production $P_{\textbf{k}c}/P_{\textbf{k}0}$ with $\tan \phi = 0.1$.
\emph{(a)} $\rho_s = 0.5$, and \emph{(b)} $\rho_s = 4$.
The black solid line denotes $E_\textbf{critical}$ for optimal $\mu_s$ for the respective density ratio $\rho_s$. The plots were obtained using standard resolvent analysis.} 
\label{fig:Figure 10}
\end{figure}

In this section, we shall look at the effect of solid density on the interaction between the compliant wall and LSMs for a loss tangent of $\tan(\phi) = 0.1$. Figure \ref{fig:10a} shows the compliant to rigid production ratio as a function of $E$ and $H$, for a density ratio $\rho_s = 0.5$. Compared to the case with $\rho_s = 1$ (Figure \ref{fig:Figure 7}b), the interaction bands have shifted towards lower $E$ regions, although the qualitative structure remains similar. This is attributed to the fact that resonant frequencies where the compliant wall response peaks scale as $\omega \approx \sqrt{E/\rho_s}$; therefore, a reduction in $\rho_s$ would require lower $E$ for matching the frequency between the flow structures and peak solid response. Due to the interaction with solids with lower $E$, the compliant surface with $\rho_s = 0.5$ responds much more strongly to the flow stresses compared to Figure \ref{fig:Figure 7}b, resulting in larger variations in $P_{\textbf{k}c}/P_{\textbf{k}0}$. The solid black line denotes the threshold modulus below which TWF instabilities are amplified. The movement of the interaction zones to lower values of $E$ results in a larger separation between interaction zones and TWF onset. Figure \ref{fig:10b} shows the production ratio contour plot for a solid with density $\rho_s = 4$. Compared to Figure \ref{fig:10a}, the interaction between solid and large scale structures occurs for solids with higher $E$, which reduces the strength of coupling between them. This results in weaker changes in resolvent gain and mode TKE production, which can be seen in the contour levels. The neutral curve for optimal damping at $\rho_s = 4$ shows minimal variation from the case of $\rho_s = 0.5$ up to $H = 0.7$, beyond which $E_\text{critical}$ increases rapidly. It can also be concluded that increasing $\rho_s$ from 0.5 to 4 not only results in a weaker coupling, but also a reduction in the width of the interaction bands across $E$. However, it also reduces the gap between the interaction regimes and instability threshold. 

\subsection{Compressible flows}

In this section, we extend the analysis performed in the previous section to viscoelastic layers interacting with supersonic turbulent boundary layers. We exclude the analysis performed for the near-wall modes (Figure \ref{fig:Figure 6}) since these modes resulted in an increase in the resolvent gain as we demonstrated in the previous section. In supersonic wall turbulence, the organization of eddies remains similar to incompressible turbulence according to Morkovin's hypothesis. DNS results \citep{pirozzoli2011turbulence, duan2011direct} at Mach numbers up to 3 also showed the similarity between streamwise and spanwise length scales for incompressible and compressible boundary layers. Hence, we use identical length and timescales to describe the VLSM mode as in incompressible flows, which is $(\lambda_x, \lambda_z, c) = (6, 1, 0.7)$. An $M_\infty = 2$ adiabatic wall turbulent boundary layer profile with $Re_\tau = 1110$ is used as the mean flow. 

\begin{figure}
    \centering
    \includegraphics[width=1\linewidth]{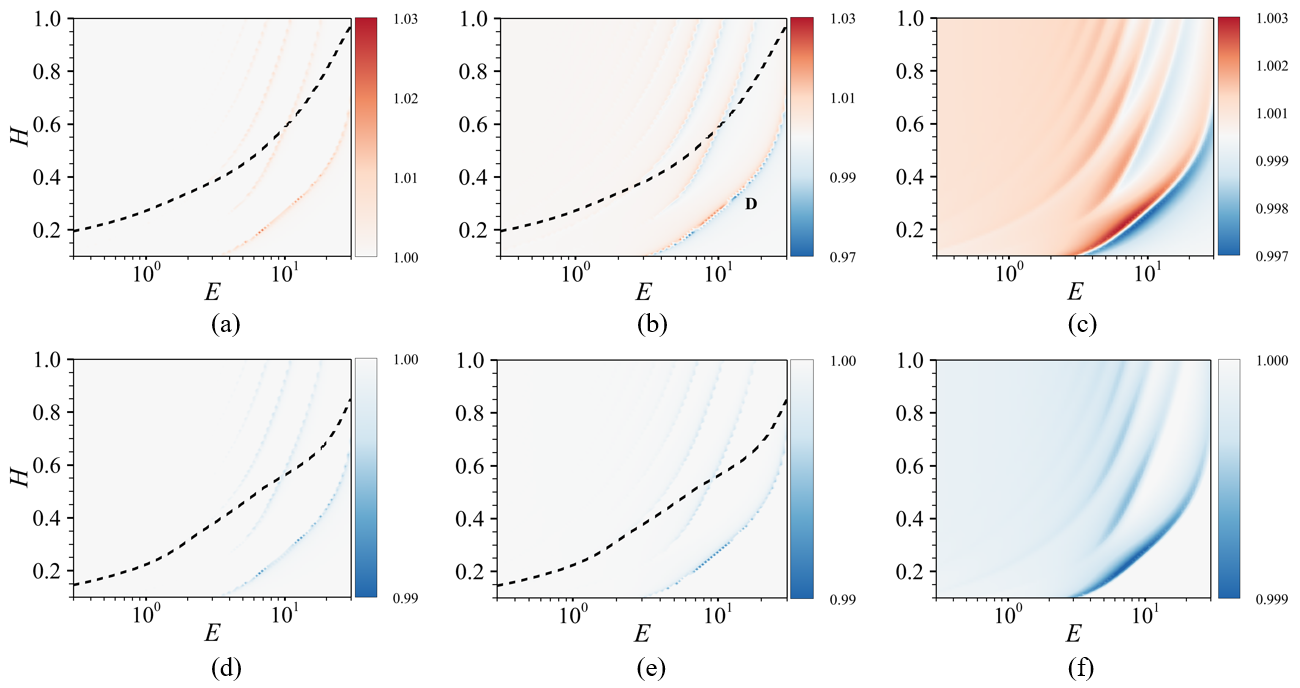}
    \caption{Contour plots of the resolvent gain and TKE production ratios using eddy viscosity \emph{(a-c)} and standard \emph{(d-f)} resolvent analysis, \emph{(a,d)} Compliant to rigid wall singular values $\sigma_{\textbf{k}c}/\sigma_{\textbf{k}0}$ for $\tan \phi = 0.01$, \emph{(b,e)} Compliant to rigid wall TKE production $P_{\textbf{k}c}/P_{\textbf{k}0}$ for $\tan \phi = 0.01$, and \emph{(c,f)} Compliant to rigid wall TKE production $P_{\textbf{k}c}/P_{\textbf{k}0}$ for $\tan \phi = 0.1$. The black dashed line denotes values of $E$ below which the viscoelastic surface becomes unstable.}
    \label{fig:Figure 11}
\end{figure}

Figure \ref{fig:Figure 11} shows the resolvent gain ratio $\sigma_{\text{k}c}/\sigma_{\text{k}0}$ and TKE production ratio $P_{\text{k}c}/P_{\text{k}0}$ in the $E-H$ plot using both standard and eddy viscosity resolvent analysis. Figure \ref{fig:Figure 11}a shows the resolvent gain ratio for materials with $\tan(\phi) = 0.01$ obtained using an eddy viscosity profile. Compared to results obtained from incompressible flows, a very weak interaction is observed, with the maximum gain increase of only $3\%$. A reduction in gain was not observed for any case, which is similar to what was observed with incompressible flow in Figure \ref{fig:Figure 7}d. The interaction occurs for materials with $(5<E<30)$, which is significantly larger than the density-matched incompressible case ($E<0.1$). This behavior follows the trends observed in the incompressible resolvent cases with increasing density ratio in Figure \ref{fig:Figure 10}. The present case at $\rho_s = 1000$ can be viewed as a progression of the trend shown with $\rho_s = 4$ in Figure \ref{fig:10b}. The shift in interaction towards higher $E$ with increasing $\rho_s$ can be explained by matching the mode frequency with characteristic structural frequencies, which can be given by $U_\infty/\delta \approx \frac{1}{H^*}\sqrt{E^*/\rho^*_s}$, which can be written as $E \approx \rho_sH^2$. Thus, for the same thickness ratio, the non-dimensional Young's modulus of interaction scales directly with the density ratio. This can be seen by the shift from $E = O(10^{-2})$ in Figure \ref{fig:Figure 7}a to $E = O(10)$ in Figure \ref{fig:Figure 11}a, where $\rho_s$ increases from 1 to 1000. Due to the large increase in $E$, the fluid-structure coupling becomes weaker. Compared to density-matched interactions, the increase in $\rho_s$ results in a decrease in the solid admittance at resonance by $\sqrt{\rho_s}$. Compared to density-matched coupling where both off resonant and resonant interactions were strong, resulting in a strong response across a continuous  range of $E$, an increase in $\rho_s$ limits these interactions to only resonant frequencies, resulting in a weak, narrow response. 

Figure \ref{fig:Figure 11}b shows the TKE production ratio [\ref{eq:29}] without pre-multiplication by $\sigma^2_\textbf{k}$ for interactions with $\tan(\phi) = 0.01$, in order to highlight the effect of mode reorganization. While the narrow banded structure remains, the interactions exhibit both reduction and amplification of TKE production by up to $3\%$. For $E<E_{\text{resonant}}$, a reduction in Reynolds stress is observed, whereas an increase is observed for $E>E_{\text{resonant}}$. While the interactions occur across multiple bands, only the first band, which occurs along the highest $E$ exhibits the strongest change in Reynolds stress. Within this band, the strongest interaction occurs between $0.2<H<0.4$. An optimal region of similar form can be seen for the density-matched cases in Figures \ref{fig:Figure 7}b and \ref{fig:Figure 7}d, where the first band shows strong coupling between $0.1<H<0.5$. This can be explained by considering the competing effects of $E$ and $H$ variation on the response of the solid. Since the viscoelastic surface is attached to a rigid base at $y = -H$, reducing $H$ reduces the solid response, whereas reducing $E$ increases the solid response. Thus, the optimal response is obtained for a specific range of $E$ and $H$, as shown in Figures \ref{fig:Figure 7}b, \ref{fig:Figure 7}d, and \ref{fig:Figure 11}b. The onset curves for flutter instabilities are shown with dashed black lines. It can be seen that while the majority of the interactions occur in the range of $E$, where the viscoelastic surface is predicted to be unstable, the first and strongest interaction band lies for values of $E$ in the stable region. This is in contrast to the case of density-matched incompressible flows where the onset of instabilities completely preceded the favorable interactions between VLSMs and the viscoelastic solid.

Figure \ref{fig:Figure 11}d shows the resolvent gain ratio calculated using standard resolvent. An increase in gain is not observed for any case, and the maximum reduction in gain is only about $1\%$. Compared to the eddy viscosity case in Figure \ref{fig:Figure 11}a, the change in resolvent gains is smaller. Figure \ref{fig:Figure 11}e shows the TKE production ratio using standard resolvent (without the pre-multiplication by $\sigma^2_\textbf{k}$). While an increase in Reynolds shear is not observed, the decrease in Reynolds shear stress is weaker than the corresponding decrease with eddy viscosity (Figure \ref{fig:Figure 11}b). Once $\sigma^2_\textbf{k}$ is included (not shown), the Reynolds shear for eddy viscosity case exhibits an overall increase across $E$ and $H$, with the maximum increase being $5\%$. For standard resolvent instead, an overall decrease is observed, with maximum reduction being $3\%$. Looking at the instability onset curve, it can be seen that removing the eddy viscosity profile results in a slightly earlier onset of instability. Thus, the influence of eddy viscosity is weakly stabilizing for the onset of flutter for a supersonic flow with high density ratio. 

Figures \ref{fig:Figure 11}c and \ref{fig:Figure 11}f show the TKE production ratio (without pre-multiplication with $\sigma^2_\textbf{k}$) for contours for compliant surfaces with higher loss tangent of $\tan(\phi) = 0.1$ using eddy viscosity and standard resolvent respectively. It is evident that an increase in damping significantly reduces the response of the viscoelastic surface, with the amplification or attenuation of Reynolds stress remaining below $1\%$. The structure of the response is similar to the low damping cases, but much wider across $E$ due to higher damping. While the eddy viscosity resolvent shows a clear change of response from decrease to increase across the resonant frequency, the standard resolvent only shows a decrease, which is also relatively weaker. However, the weak response of the solid to flow perturbations makes these interactions unlikely to  influence drag.

\begin{figure}
   \centering
    	\begin{subfigure}[t]{0.48\textwidth}
		\includegraphics[width=\textwidth]{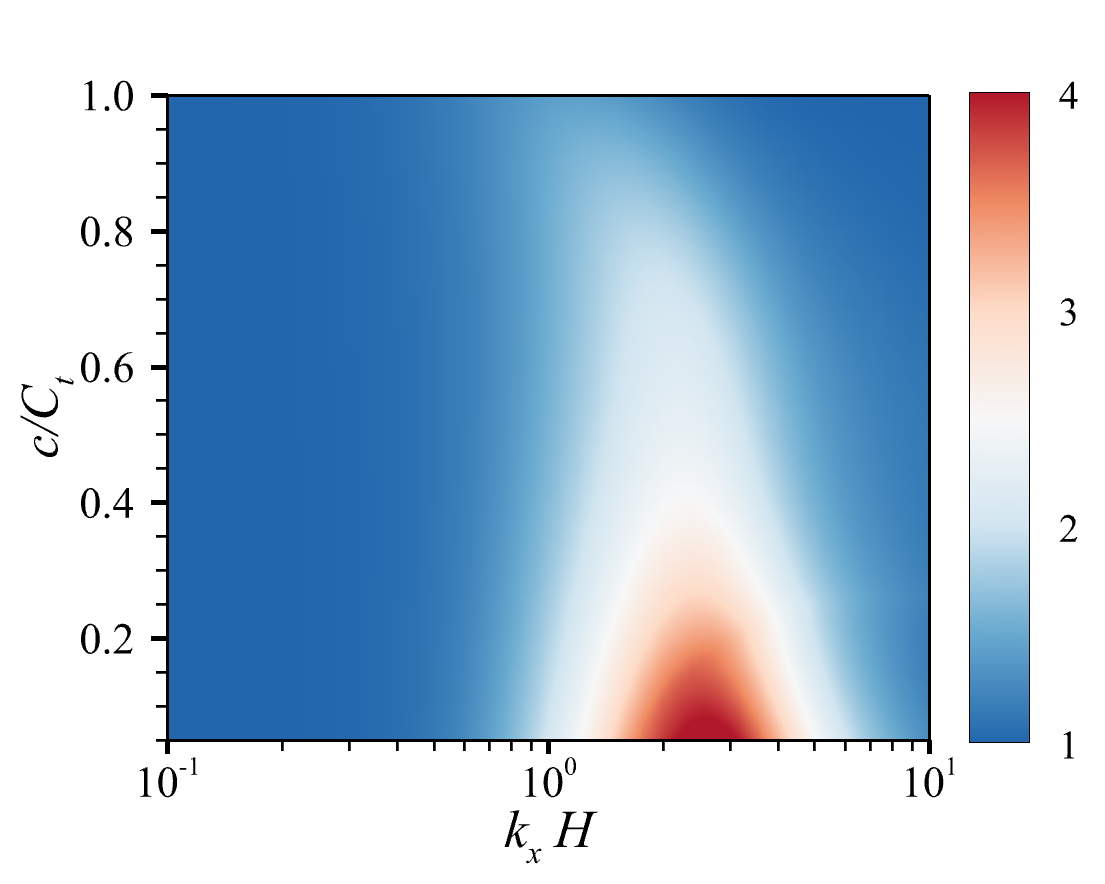}
		\caption{}
		\label{fig:12a}
	\end{subfigure}
	\hfill
	\begin{subfigure}[t]{0.48\textwidth}
		\includegraphics[width=\textwidth]{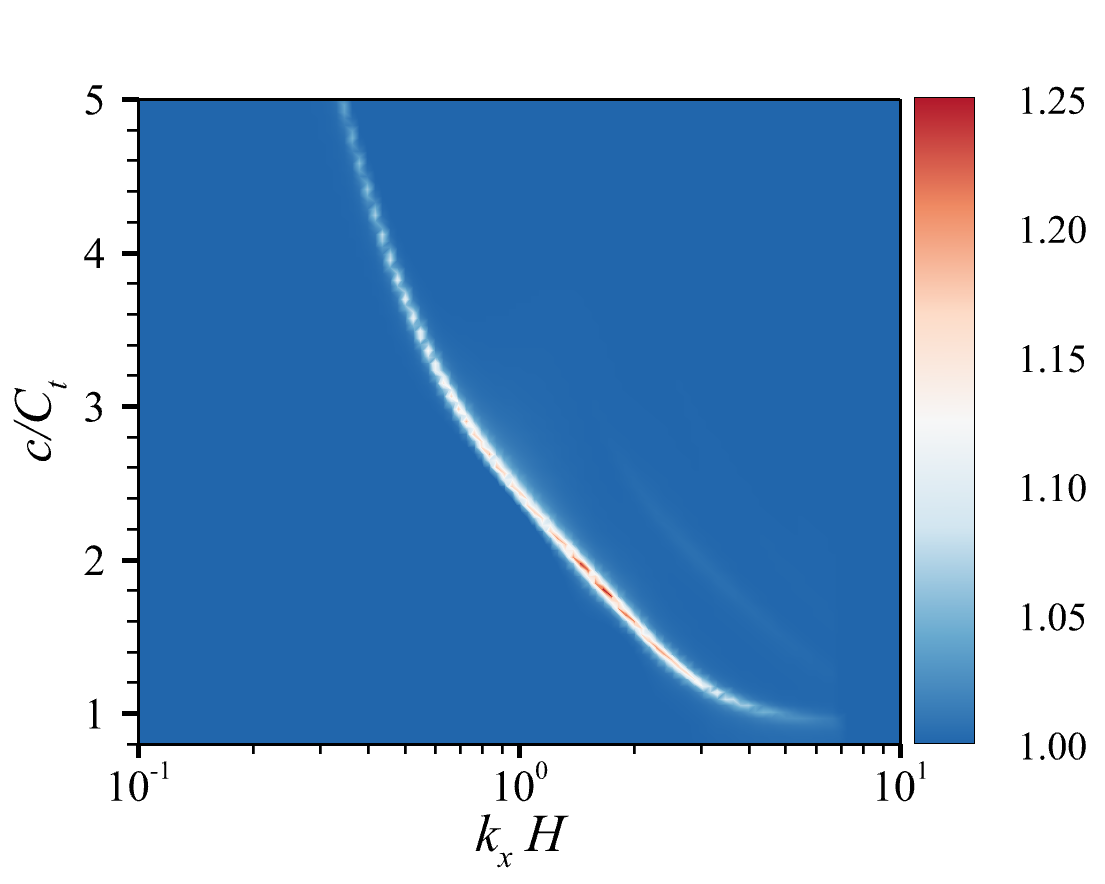}
		\caption{}
		\label{fig:12b} 
	\end{subfigure}	
        \hfill
\caption{Resolvent gain ratio $\sigma_{\text{c}}/\sigma_{0}$ across $k_x$ and $c$ for cases \emph{(a)} $A$ in Figure \ref{fig:Figure 7}c, and \emph{(b)} $D$ in Figure \ref{fig:Figure 11}b. The wavenumber and phase speeds are normalized by the solid quantities.} 
    \label{fig:Figure 12}
\end{figure}

While we see in Figure \ref{fig:Figure 11}b that materials in the first interaction band do exist in the stable regime, the non-normality of the Navier-Stokes operator can cause large amplification of fluid-structure modes even for materials that can be linearly stable, as observed \citep{pfister2022global, luhar2015framework, tsigklifis2017interaction} for incompressible flows. 
Figure \ref{fig:Figure 12} shows the resolvent gain ratio $\sigma_c/\sigma_0$ of two-dimensional modes $(k_z = 0)$ in the $k_x - c$ plane for specific viscoelastic solids $A$ (Figure \ref{fig:Figure 7}c), and $D$ (Figure \ref{fig:Figure 11}b) respectively. The wavenumber $k_x$ and wave-speed $c$ are normalized by the solid thickness $H$ and shear speed $C_t$,
respectively. For highly damped viscoelastic solids such as $A$, static divergence is the primary mode of instability. For $2<k_x H<3$, Figure \ref{fig:12a} shows strong amplification of up to four times at near zero wave-speeds, which is a known characteristic of this instability \citep{gad1984interaction, yeo2001turbulent}. Despite the amplification, the absolute resolvent gain of the amplified divergence mode for $A$ is $\sigma_{\text{SD}} \approx 3$ and the absolute gain of the VLSM modes is $\sigma_\text{VLSM} \approx 220$. This implies that the divergence amplification may not precede the favorable interactions described in Figure \ref{fig:Figure 7}c. 

Figure \ref{fig:12b} shows the resolvent gain ratio for the material $D$ indicated in Figure \ref{fig:Figure 11}b across the $k_x - c$ space. The spanwise wavenumber is set to $k_z = 0$. The resolvent gain ratio is larger than unity for a streak extending across $k_x$ and $c$. This is one of the natural modes of the viscoelastic coating. The peak amplification of $25\%$ occurs near $1<k_xH <2$ and $1.5<c/C_t<2$. Since these modes are not characteristically high gain resolvent modes, this amplification is unlikely to cause earlier onset to flutter.

\begin{figure}
   \centering
    	\begin{subfigure}[t]{0.48\textwidth}
		\includegraphics[width=\textwidth]{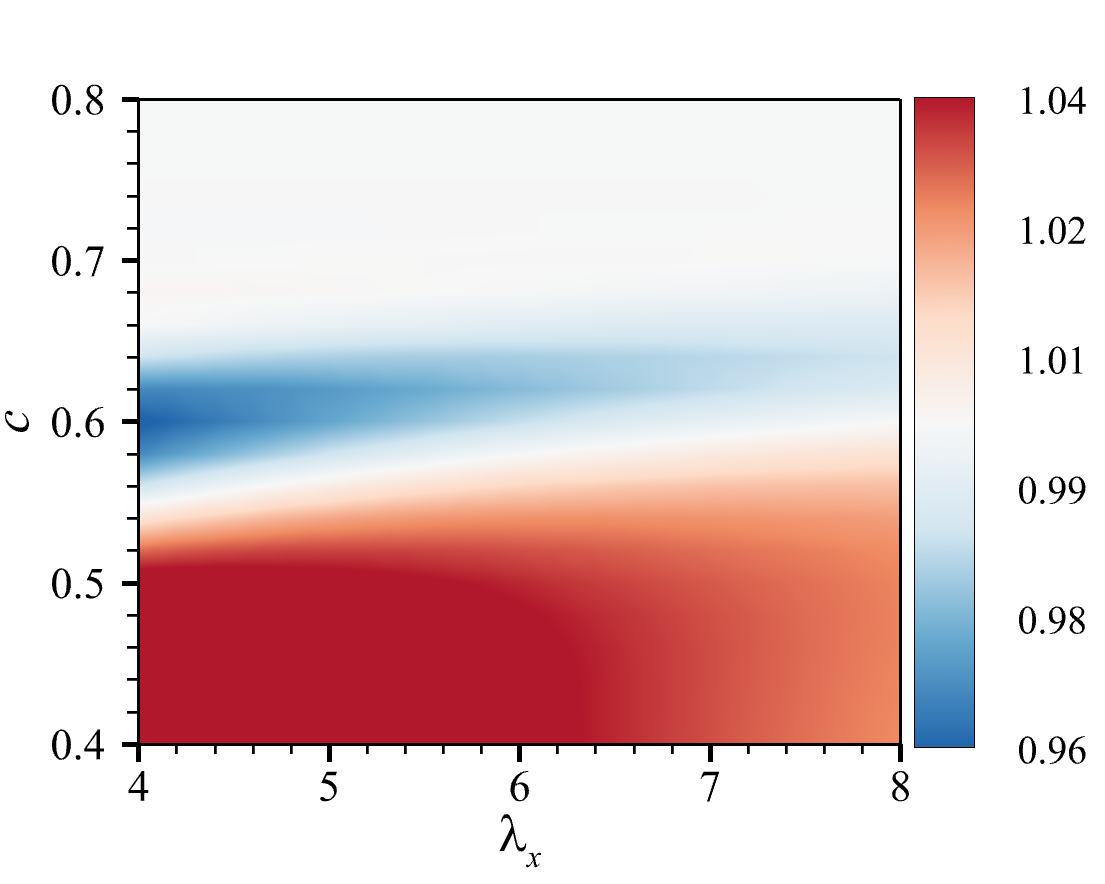}
		\caption{}
		\label{fig:13a}
	\end{subfigure}
	\hfill
	\begin{subfigure}[t]{0.48\textwidth}
		\includegraphics[width=\textwidth]{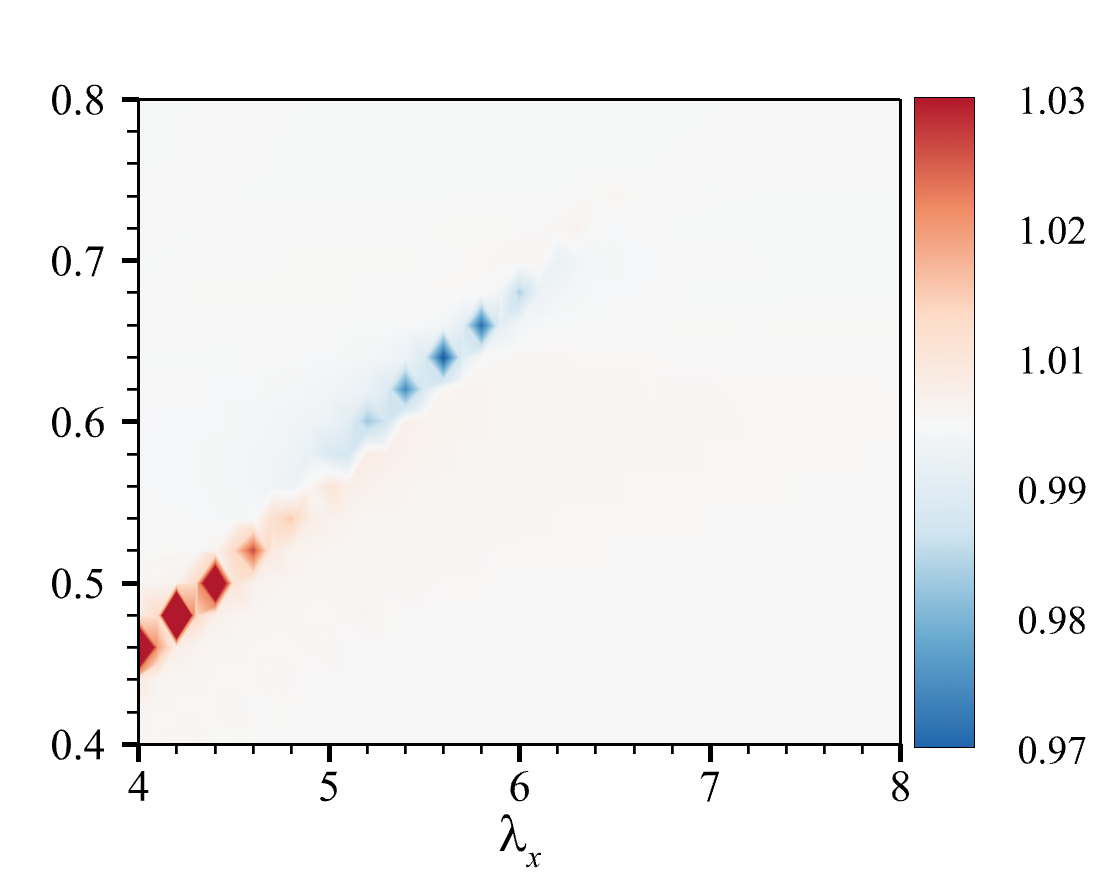}
		\caption{}
		\label{fig:13b} 
	\end{subfigure}	
        \hfill
\caption{TKE production ratio across $\lambda_x$ and $c$ for cases \emph{(a)} $A$ in Figure \ref{fig:Figure 7}c, and \emph{(b)} $D$ in Figure \ref{fig:Figure 11}b. The spanwise wavelength is set to $\lambda_z = 1$. The wavelengths and phase speeds are normalized by fluid quantities. The calculations were performed using standard resolvent.} 
    \label{fig:Figure 13}
\end{figure}

So far, we have demonstrated how the material properties of a viscoelastic layer influences its interaction with a single large scale mode. Figure \ref{fig:Figure 7}c and \ref{fig:Figure 11}b has shown how certain fluid-solid coupling in the linearly stable regime can weakly suppress the Reynolds shear stress associated with a single large scale mode. However, the Reynolds shear stress is an integral of the contribution over all scales. Figure \ref{fig:Figure 14} shows the integrated Reynolds shear component for cases $A$ and $D$ in Figures \ref{fig:Figure 7}c and \ref{fig:Figure 11}b respectively. Case A shows that suppression peaks for mode speeds $0.55<c<0.65$ and towards smaller wavelengths. With increasing $\lambda_x$, the reduction in Reynolds stress becomes weaker. However, for $c<0.5$, an amplification of Reynolds stress is observed over all scales, with the maximum occurring for smaller wavelengths. Faster scales with $c>0.7$ are not found to interact with the viscoelastic surface. For case $D$, the stronger interactions that suppress the Reynolds stress can be found along a straight line ranging between $5<\lambda_x<6$ and $0.6<c<0.7$. However, interactions with. For both case $A$ and $D$, the amplification of Reynolds stress by lower speed modes slightly exceeds the suppression observed for faster modes. This could be related to the proximity of the modes to wall, where slower modes lying closer to wall can interact relatively strongly compared to faster modes lying further away. 

\subsection{Admittance of viscoelastic layers}

\begin{figure}
	\centering
    	\begin{subfigure}[t]{0.31\textwidth}
		\includegraphics[width=\textwidth]{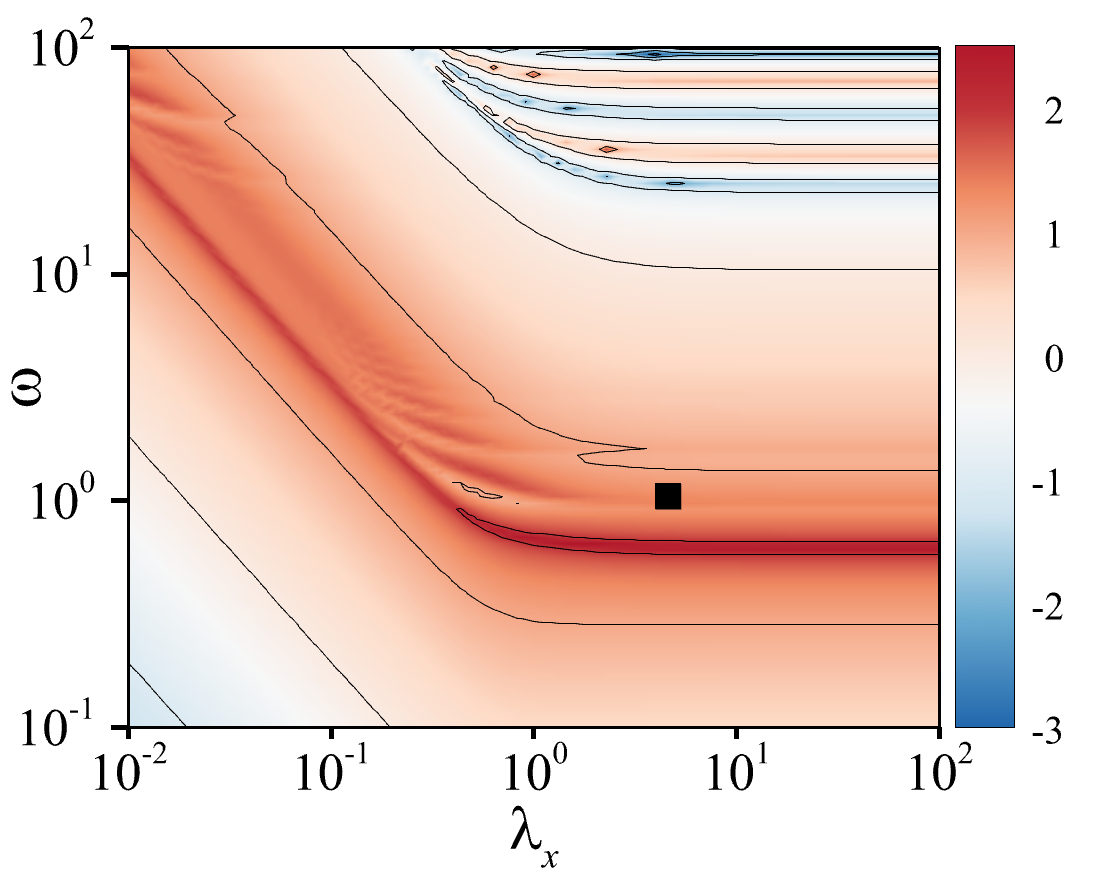}
		\caption{}
		\label{fig:14a}
	\end{subfigure}
	\hfill
	\begin{subfigure}[t]{0.31\textwidth}
		\includegraphics[width=\textwidth]{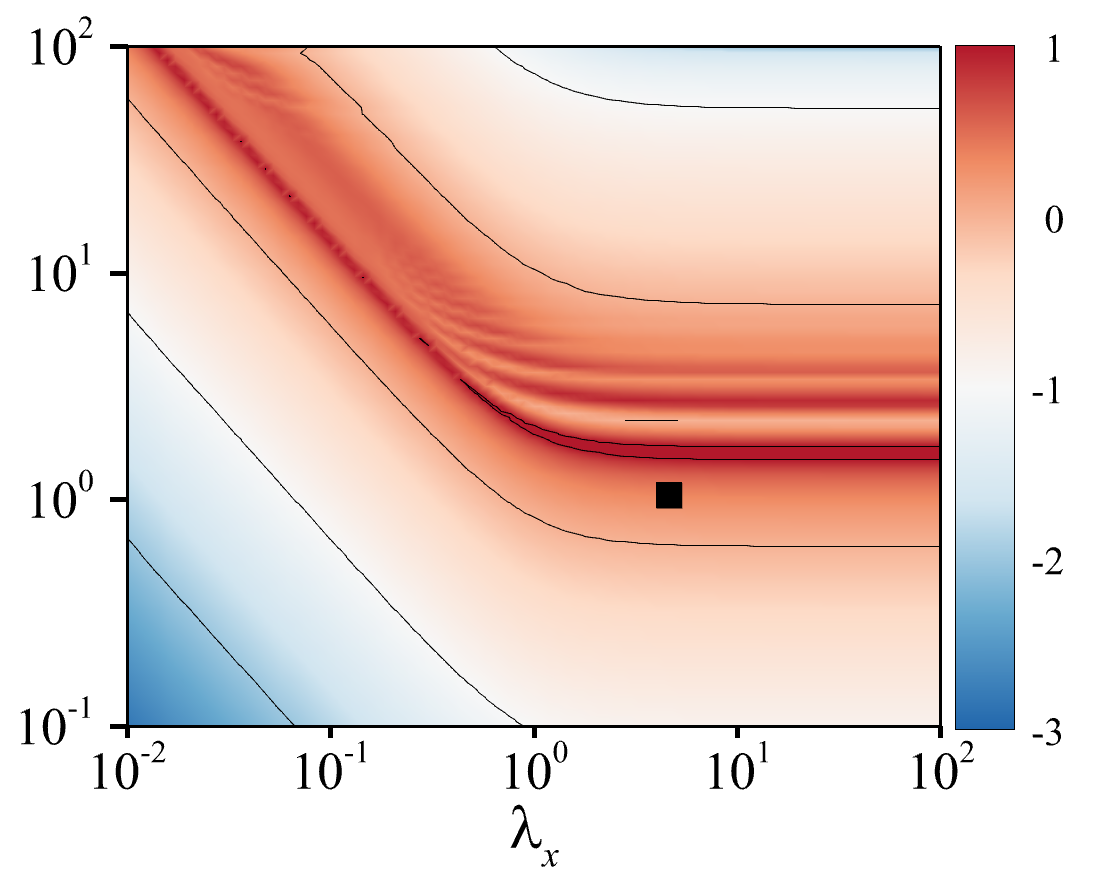}
		\caption{}
		\label{fig:14b} 
	\end{subfigure}	
        \hfill
        \begin{subfigure}[t]{0.31\textwidth}
		\includegraphics[width=\textwidth]{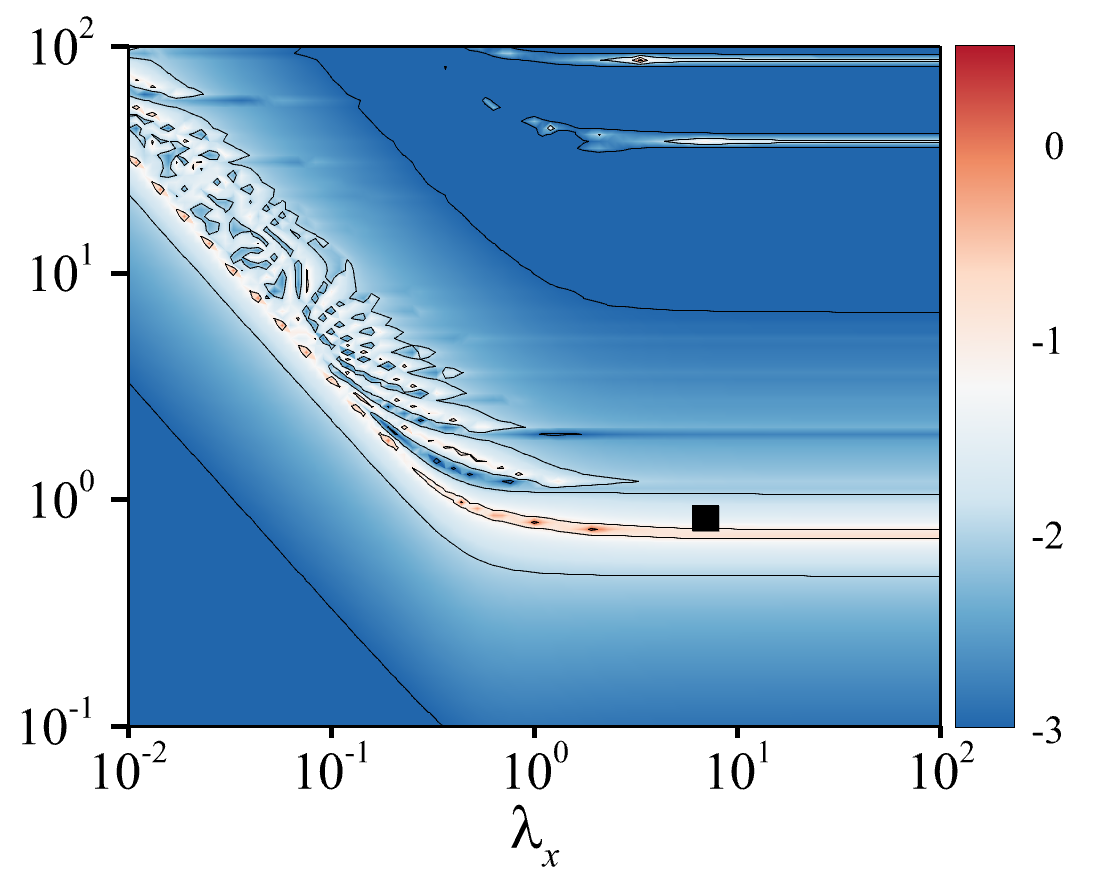}
		\caption{}
		\label{fig:14c} 
	\end{subfigure}	
        \hfill

\caption{Contour plots showing absolute value of admittance $\log_{10}{(Y_{22})}$ for highlighted materials \emph{(a)} Case $C$ with $\rho_s = 0.5$, \emph{(b)} Case $B$ with $\rho_s = 1$, and \emph{(c)} Case $D$ with $\rho_s = 1000$. The spectral location of LSMs is marked as $\lambda_x = 4, \omega = 0.9$ for \emph{(a,b)}, and $\lambda_x = 6, \omega = 0.7$ for \emph{(c)}.} 
\label{fig:Figure 14}
\end{figure}

Figure \ref{fig:Figure 7} and \ref{fig:Figure 11} show a vast range of interactions that are theoretically possible between viscoelastic coatings with different properties and high gain turbulent structures like VLSMs. In this section we shall look into the response of the highlighted materials $B,C$, and $D$ from Figure \ref{fig:Figure 7}e, \ref{fig:10a}, and \ref{fig:Figure 11}b respectively in the spectral space. Particularly, we shall look into the admittance spectrum $Y_{ij}(\textbf{k)}$ of these materials, which is defined as the velocity of the surface per unit surface traction. For a continuum viscoelastic surface, the admittance is a second-order tensor in 3D. In this case, we shall look into the tensor components $Y_{22}$, which provides the wall-normal velocities of the compliant surface in response to fluctuating normal stress on the flow facing side of the viscoelastic surface. Figure \ref{fig:Figure 14}(a,b) shows the magnitude of $Y_{22}$ as a function of $\omega$ and $\lambda_x$ for materials $C$ and $B$ respectively, as marked in Figures \ref{fig:10a} and \ref{fig:Figure 7}e. The spanwise wavelength is fixed at $\lambda_z = 1$, corresponding to the large scale motions. The admittance values are normalized by the freestream fluid variables.

The black highlighted points denote the spectral location of the VLSM mode given by $(\lambda_x,\omega) = (4,0.92)$. Figure \ref{fig:14a} shows the admittance spectrum for $C$, which has density ratio $\rho_s = 0.5$. For the viscoelastic solid to interact strongly with the VLSM mode, the mode must spectrally lie within regions where the $Y_{22}$ is $O(1)$ or higher. It can be seen from the contour lines that the peak admittance is more than 100, and the regions with $Y_{22}>O(1)$ occupy a significant region in the spectral space. With increase or decrease in $E$, the admittance spectrum would move towards higher or lower frequencies respectively, while the VLSM mode remains fixed. With such a wide region with $Y_{22}>O(1)$, a large range of materials with different $E$ can interact strongly, as was observed in Figure \ref{fig:10a}. For $B$ in Figure \ref{fig:14b}, the density ratio is $\rho_s = 1$. The peak admittance is now reduced by an order of magnitude compared to $\rho_s = 0.5$, and moreover, the $\omega$ bandwidth where $Y_{22}>O(1)$ has shrunk significantly, thus restricting the interactions to a smaller range of $E$, as observed in Figure \ref{fig:Figure 7}(b,e).   

Finally, Figure \ref{fig:14c} shows the $Y_{22}$ spectrum for $D$ which has a density ratio $\rho_s = 1000$, and the VLSM frequency is marked as $(\lambda_x,\omega) = (6,0.7)$. It can be seen that the peak admittance is just about $O(1)$ in a few locations, and the peak admittance reached within the resonance band is about 0.1, which also has a narrow $\omega$ bandwidth. Thus, interactions are expected to grow weaker and become limited to a small range of $E$, as shown in Figure \ref{fig:Figure 11}. Thus, for low $\rho_s$, both resonant and off-resonant interactions are strong and may give rise to favorable changes in resolvent gain or Reynolds stress, whereas only resonant interactions are viable at high density ratios.

\section{Conclusions}

In this work, we use resolvent analysis to study the interaction between an isotropic viscoelastic compliant surface and a turbulent boundary layer to understand its potential for mitigating specific turbulent scales responsible for drag. Since compliant walls are known to undergo fluid-structure instabilities for certain choices of solid parameters, the resolvent analysis is complemented with linear stability analysis to segregate the stable interactions. We perform a systematic study over all non-dimensional solid parameters such as Young's modulus $E$, layer thickness $H$, solid viscosity $\mu_s$ and solid density $\rho_s$ across both incompressible and compressible regimes. We also include an eddy viscosity profile modeled using the mixing length model, and compare results from both standard and eddy viscosity resolvent analysis. 

For density-matched incompressible flows, the low rank maps in spectral space show that the canonical high gain resolvent modes such as the near-wall cycle and large scale structures remain low-rank on the inclusion of a viscoelastic surface, thereby enabling the study of interactions on a mode by mode basis using the leading resolvent mode. However, distinct low rank bands emerge on the inclusion of the compliant surface, which occur for large spanwise wavelengths $(\lambda_z^+>1000)$, and have a constant streamwise wavelength. For a fixed choice of mode speed $c$, the band wavelength $\lambda_x$ scales with the solid thickness $H$. 

Based on the low rank maps, two wavenumber frequency triplets were identified as proxies for near-wall cycle and large scale structures. A sweep over modulus $E$ and thickness $H$ showed that the near-wall modes resulted in an increase in resolvent gain and integrated Reynolds shear stress across all possible combinations, consistent with previous studies \citep{luhar2015framework, song2026structured}. For the large scale modes, sweeps were performed across $E$ and $H$ for materials with low and high viscosity $\mu_s$. When an eddy viscosity profile was not included, the majority of materials with modulus below $E<0.1$ and low damping exhibited a reduction in resolvent gain and integrated Reynolds shear stress by up to $50\%$. However, the neutral stability curves showed that these favorable interactions would be preceded much earlier by traveling wave flutter, rendering these interactions impractical. Increasing the $\mu_s$ spread the reduction in Reynolds stress to materials with $E \approx 1$, but reduced the magnitude of attenuation possible. The higher $\mu_s$ delayed the onset of instability, which enabled certain sub-optimal interactions (reductions below $5\%$) with large $H$ to extend to regions beyond the neutral curves ($E>E_\text{neutral}$). Inclusion of an eddy viscosity profile resulted in regions in the $E-H$ space where both increase and decrease of Reynolds shear stress was observed to a much higher degree for low $\mu_s$. However, the interactions still remained forbidden due to the prior onset of flutter. In this case, an increase in $\mu_s$ did not result in a spread of the favorable interaction regions beyond the stability margin. Since interactions with highly damped materials showed a favorable response, an attempt was made to delay the onset of flutter as much as possible to extract stronger interactions. An optimal $\mu_s$ was found at the intersection of traveling wave flutter and divergence instabilities for different values of $H$, where $E_\text{neutral}$ was minimum. However, the difference between this value and the asymptotic $E_\text{neutral}$ at high damping was not high enough to significantly enhance the range of interactions. 

Studying the effect of density ratio $\rho_s$ on the interaction between compliant surfaces and large scale modes shows that increasing $\rho_s$ shifts interaction towards higher modulus $E$, thereby weakening it, while reducing $\rho_s$ does the opposite. Moving onto compressible flows at $M_\infty = 2$ and $\rho_s = 1000$, the effect of $\rho_s$ is realized much more strongly. Large $\rho_s$ shifts the interactions to materials with $E \approx 10$, which is much higher than the instability threshold. However, the weakening of the interactions results in a maximum Reynolds stress reduction of up to $3\%$, which is again realized for very lightly damped materials. Increasing the damping by an order of magnitude reduces this attenuation to below $1\%$. The admittance contours of favorably interacting viscoelastic materials with different values of $\rho_s$ show that low $\rho_s$ enables strong interactions between viscoelastic surfaces and flow structures for a wide frequency band around resonance, resulting in a broader range of coupling across materials with different $E$, $H$ and $\mu_s$. Increasing $\rho_s$ to values similar to supersonic flows results in a decrease in admittance, with weaker interactions only along narrow resonant frequencies. While this constrains the range of viscoelastic materials which can be used to design drag reduction strategies, it shifts these interactions far into the stable regime with relatively small non-normal amplification.

%Moving forward, future research should be directed towards improving the accuracy of the current models. Real viscoelastic materials exhibit frequency dependent storage and loss modulus, which can affect the flutter thresholds and favorable interaction regimes. Given that the favorable regimes identified in the current work lie close to the onset of flutter, incorporation of frequency dependent effects can vastly improve the prediction of the physics.  In this study we used a fixed eddy viscosity model, which is known to overestimate the energy transfer near the wall \citep{symon2023use}, resulting in excessively strong interactions between the selected scales and compliant wall. Optimization of the eddy viscosity profile or using scale dependent eddy viscosity may improve the accuracy of the models. In the current work, we show for incompressible flows that materials which favorably modulate large scale structures are at least 10 times softer than the flutter threshold. Although implemented to suppress transient growth, recent application \citep{fabbiane2025phononic} of Bragg scattering band gaps to attenuate TWF for laminar flows can be used improve the margin for favorable interactions in the current setup, however, delaying TWF significantly beyond the linear stability threshold may remain challenging. 

\section{Acknowledgments}
	
The work was supported by grant number N00014-21-1-2005 from the Office of Naval Research with Dr. Leighton Myers as the program monitor,  and US Army Research Office grant number W911NF261A244 with Dr. Kenneth Granlund as the Program Officer. 
	
%\section*{Appendix}

\appendix

\section{Comparison of standard and eddy resolvent modes}

Figure \ref{fig:Figure 7} reports strong differences in the amplification behavior of the coupled fluid-structure modes on the inclusion of an eddy viscosity profile. At high Reynolds numbers, since wall pressure fluctuations are the primary driver of these interactions, a component-wise budget analysis of the pressure can provide insights on these differences. On taking the divergence of the momentum equations, the pressure Poisson equation for a triplet $\textbf{k}$ becomes

\begin{equation}
    \nabla^2p_{\textbf{k}} = -2ik_xv_{\textbf{k}}\frac{d\bar{U}}{dy} + \frac{2}{Re}\frac{d \nu_t}{dy}\nabla^2 v_\textbf{k} +\frac{2}{Re}\frac{d^2\nu_t}{dy^2}\frac{dv_\textbf{k}}{dy} + \nabla . f_\textbf{k}, 
    \label{eq:30}
\end{equation}

\begin{equation}
    \left.\frac{d p_\textbf{k}}{dy}\right|_{y = 0} = \frac{1}{Re}\nabla^2 v_{\textbf{k},{y = 0}}
    \label{eq:31}
\end{equation}

\begin{figure}
	\centering
    	\begin{subfigure}[t]{0.31\textwidth}
		\includegraphics[width=\textwidth]{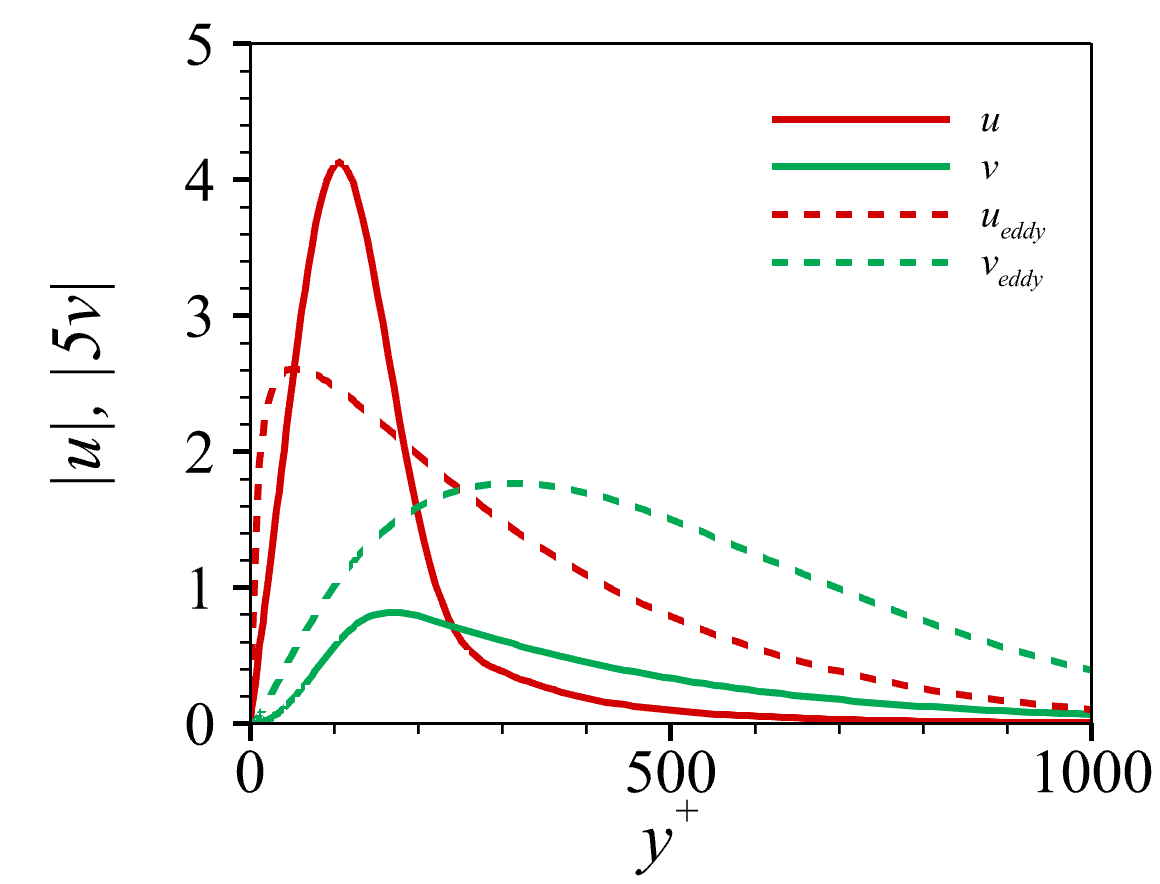}
		\caption{}
		\label{fig:15a}
	\end{subfigure}
	\hfill
	\begin{subfigure}[t]{0.31\textwidth}
		\includegraphics[width=\textwidth]{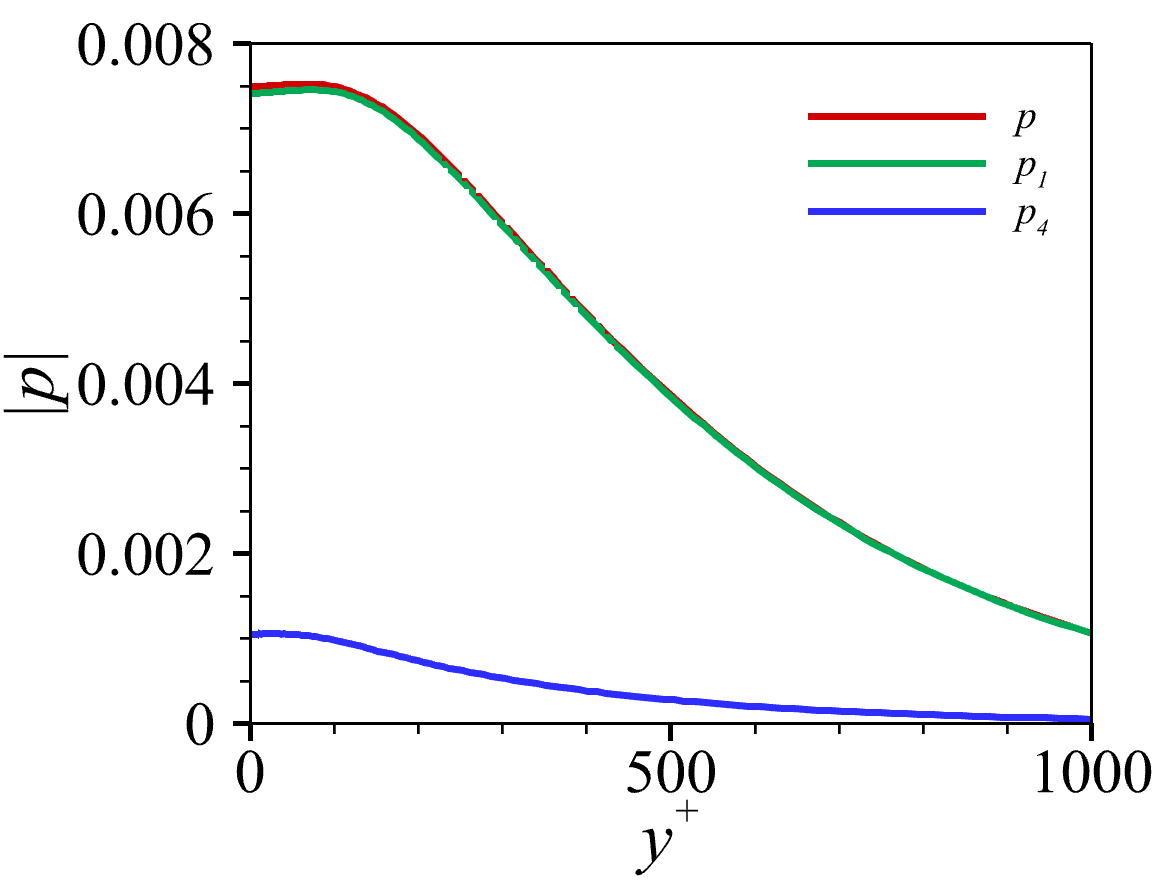}
		\caption{}
		\label{fig:15b} 
	\end{subfigure}	
        \hfill
        \begin{subfigure}[t]{0.31\textwidth}
		\includegraphics[width=\textwidth]{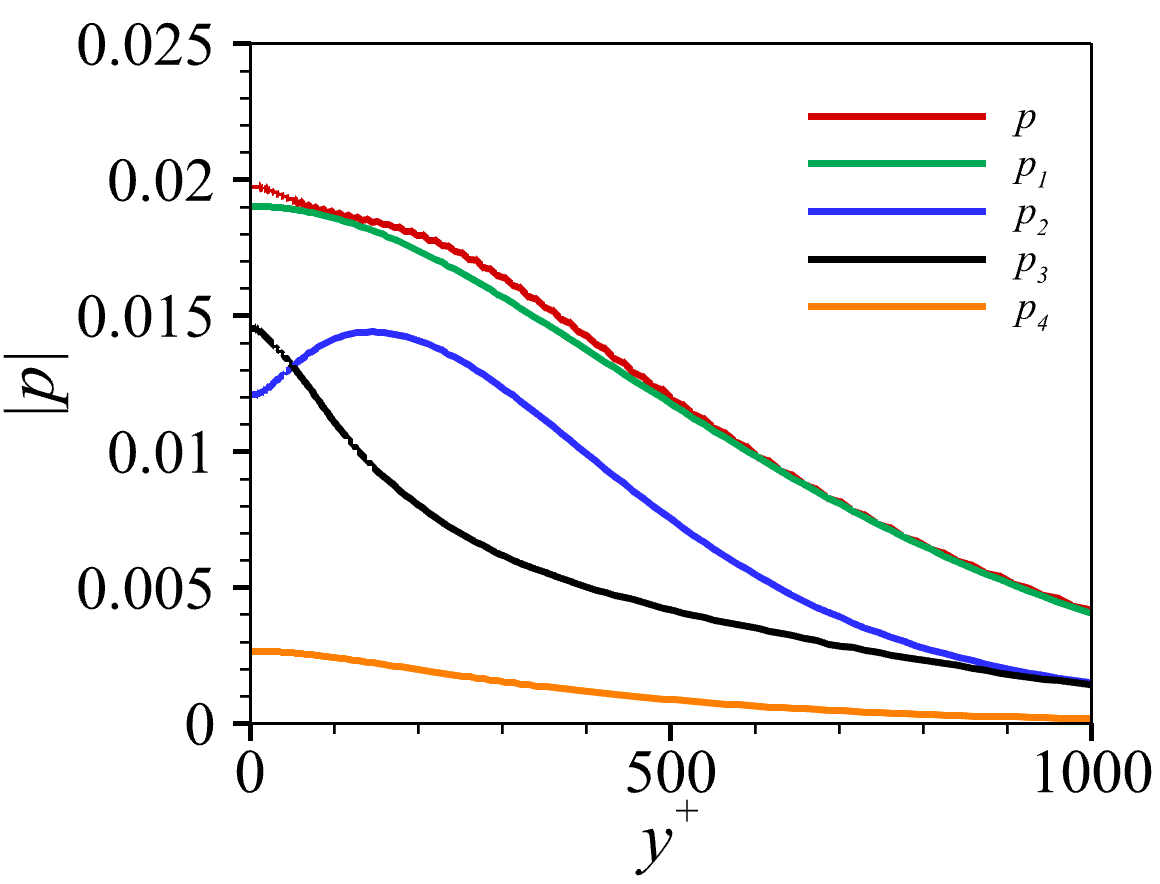}
		\caption{}
		\label{fig:15c} 
	\end{subfigure}	
        \hfill

    %\pdfcomment{what do negative values mean?}
\caption{Difference between standard and eddy resolvent modes for $(\lambda_x,\lambda_z,c)$ = (4,1,0.6) \emph{(a)} Velocity response modes, \emph{(b)} Pressure source terms for standard resolvent, and \emph{(c)} Pressure source terms for eddy resolvent. } 
\label{fig:Figure 15}
\end{figure}

Using [\ref{eq:30}], the pressure produced by each source term can be computed separately assuming homogeneous Neumann boundary conditions at $y = 0$ and homogeneous Dirichlet conditions at the freestream boundary. The pressure component that arises due to the non-homogeneous boundary condition in [\ref{eq:31}] is small at high Reynolds numbers \citep{luhar2014structure}. For standard resolvent where eddy viscosity $\nu_t$ is zero, the second and fourth terms in the right-hand side vanish, leaving behind the form described by \cite{luhar2014structure}. The first (linear) source term gives rise to the fast pressure component, and the last (nonlinear interactions) term gives rise to the slow component. Figure \ref{fig:15b} and \ref{fig:15c} show the individual pressure contribution from each source term for standard and eddy viscosity resolvent respectively, along with the overall pressure obtained from resolvent calculation. The modes are calculated for a rigid wall case at $(\lambda_x,\lambda_z,c) = (4,1,0.6)$, which corresponds to the large scale modes investigated in Figure \ref{fig:Figure 7}. For standard resolvent, it can be seen that the fast pressure term obtained from linear interactions heavily dominates the total pressure, as has been demonstrated by \cite{luhar2014structure}. For the eddy resolvent, the fast pressure term is still seen to dominate the total pressure with minor differences near the wall. The components arising from the eddy viscosity based source terms also have relatively high magnitudes, but they have a phase difference of $\pi$ between them, which largely cancels their contribution. The nonlinear slow pressure term $p_4$ contributes the smallest in both cases. Now, looking at the magnitude of the overall pressure, the pressure from eddy resolvent exhibits three times higher magnitude at the wall compared to the standard resolvent, and this difference can be almost completely attributed to the first source term, which involves the interaction between wall-normal velocity fluctuations and the mean velocity gradient. Figure \ref{fig:15a} shows the velocity mode shapes for both standard and eddy resolvent. Compared to standard resolvent, the eddy modes are less anisotropic and have broader support along the wall-normal direction. This results in a much broader and larger magnitude $v$ response for the eddy mode at all wall-normal locations, as can be seen in Figure \ref{fig:15a}. When multiplied with mean shear, this creates a stronger source term for the pressure Poisson equation in [\ref{eq:30}], which is the likely reason for the higher wall pressure obtained for eddy resolvent modes, and eventually a stronger fluid-structure coupling.

\bibliographystyle{jfm}
\bibliography{jfm}

\end{document}